\documentclass[11pt,a4paper]{article}
\usepackage[utf8]{inputenc}
\usepackage{faktor}
\usepackage[english]{babel}
\usepackage[dvipsnames]{xcolor}
\usepackage{amsmath,amssymb,fullpage,amsthm}
\usepackage{thmtools,thm-restate}
\usepackage{ifthen,graphics}
\usepackage{xspace}
\usepackage[normalem]{ulem}
\usepackage[backend=biber,style=numeric,maxbibnames=99,citetracker=true,url=false,doi=false,giveninits=true,sortcites]{biblatex}
\usepackage[vlined,linesnumbered,boxed]{algorithm2e}
\usepackage{hyperref}
\usepackage[noabbrev,capitalise]{cleveref}
\usepackage{enumerate}
\usepackage{dsfont}
\usepackage{xspace}
\usepackage[caption=false]{subfig}

\newcommand{\orcidlogo}{\includegraphics[height=\fontcharht\font`A]{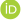}}
\newcommand{\orcid}[1]{\,\textsuperscript{%
    \href{https://orcid.org/#1}{\sf\orcidlogo\hspace{0.25em}#1}%
  }%
}%

\usepackage{tikz}
\usetikzlibrary{arrows,shapes,automata,petri,positioning}

\tikzset{
    place/.style={
        circle,
        draw=black,
        minimum size=5mm,
    }
}

\DeclareNameAlias{default}{family-given}

\SetCommentSty{textit}
\SetKwComment{Comment}{$\rhd$}{}

\SetArgSty{}
\SetKwHangingKw{KwInit}{Init:}
\SetKwProg{WhenDo}{when}{ do}{}
\SetKwProg{Function}{function}{ is}{}
\SetAlgoVlined
\DontPrintSemicolon

\newcommand{\AlgoSkip}{\vspace{0.5em}}

\crefname{equation}{Eq.}{Eqs.}
\crefname{line}{line}{lines}

\DeclareMathOperator{\Exists}{\exists\,}

\usepackage{tikz}
\tikzstyle{myRec}=[draw,rectangle,minimum height=0.22cm]
\tikzstyle{linearPoint}=[draw,rectangle,minimum height=0.32cm,minimum
width=0.05cm,inner sep=0cm,fill=black]

\newcounter{linecounter}
\newcommand{\linenumbering}{\ifthenelse{\value{linecounter}<10}{(\arabic{linecounter})}{(\arabic{linecounter})}}

\renewcommand{\thelinecounter}{\ifnum \value{linecounter} > 9\else \fi
	\arabic{linecounter}}

\SetKw{wwait}{wait}
\SetKw{Raise}{raise}
\SetKw{Return}{return}

\newcommand{\del}{{\mathit{del}}}
\newcommand{\sn}{{\mathit{sn}}}
\newcommand{\statevar}{{\mathit{state}}}

\newcommand{\op}{\textsf{op}}
\newcommand{\update}{\textsf{up}}
\newcommand{\outputFunc}{{\sf output}}
\newcommand{\val}{\mathit{val}}

\newcommand{\unknown}{{?}_{\mathbb{V}}}

\newcommand{\abort}{{\tt{abort}}}

\newcommand{\brdeliveredtext}{br-delivered\xspace}

\newcommand{\rbroadcast}{{\sf{r\_broadcast}}}

\newcommand{\rdelivered}{{\sf{r\_delivered}}}

\newcommand{\proj}[2]{\mathsf{proj}_{#2}(#1)}

\newcommand{\multip}[1]{\textbf{1}_{#1}}

\newcommand{\cat}{\oplus}

\newcommand{\valid}{\ensuremath{\mathsf{valid}}\xspace}

\newcommand{\isdefinedas}{~~\stackrel{\rm def}=~~}

\makeatletter 
\newcommand\mynobreakpar{\nobreak\@afterheading} 
\makeatother

\newtheorem{definition}{Definition}
\newtheorem{theorem}{Theorem}
\newtheorem{lemma}{Lemma}
\newtheorem{corollary}{Corollary}

\newcommand{\toto}{xxx}
\newenvironment{proofT}{\noindent{\bf
		Proof }\mynobreakpar} {\hspace*{\fill}$\Box_{Theorem~\ref{\toto}}$\par\vspace{3mm}}
\newenvironment{proofL}{\noindent{\bf
		Proof }} {\hspace*{\fill}$\Box_{Lemma~\ref{\toto}}$\par\vspace{3mm}}
\newenvironment{proofC}{\noindent{\bf
		Proof }} {\hspace*{\fill}$\Box_{Corollary~\ref{\toto}}$\par\vspace{3mm}}

\title{\bf{Process-Commutative Distributed Objects:\\
From Cryptocurrencies\\
to Byzantine-Fault-Tolerant CRDTs}}

\author{%
  \parbox{0.8\linewidth}{\centering%
    Davide~Frey$^\dagger$\orcid{0000-0002-6730-5744},
    Lucie~Guillou$^\star$,
    Michel~Raynal$^\dagger$\orcid{0000-0002-3355-8719},
    \underline{François Taïani}$^{\dagger,}$%
    \footnote{Corresponding author: François Taïani, \href{mailto:francois.taiani@irisa.fr}{\tt francois.taiani@irisa.fr},
    Authors' email addresses:
   {\tt%
    davide.frey@inria.fr, guillou@irif.fr, \{michel.raynal, francois.taiani\}@irisa.fr%
   }
    }\kern 0.5em%
    \orcid{0000-0002-9692-5678}%
  }\\
  \emph{$^\dagger$Univ Rennes, Inria, CNRS, IRISA, France,}\\ \emph{$^\star$IRIF, CNRS, Université Paris Cité, France}\\
}

\bibliography{FIFO_Byz_Traces}

\date{}
\begin{document}
\maketitle
\begin{abstract}
  This paper explores the territory that lies between best-effort Byzantine-Fault-Tolerant Conflict-free Replicated Data Types (BFT CRDTs) and totally ordered distributed ledgers, such as those implemented by Blockchains. 
  It formally characterizes a novel class of distributed
  objects that only requires a First In First Out (FIFO) order on the object operations from each process (taken individually). The formalization leverages {\it Mazurkiewicz traces} to define legal sequences of operations and ensure both Strong Eventual Consistency (SEC) and Pipleline Consistency (PC).
  The paper presents a generic algorithm that implements this novel class of distributed
  objects both in a crash- and Byzantine setting. It also illustrates the practical interest of the proposed approach using four instances of this class of objects, namely money transfer, Petri nets, multi-sets, and concurrent work stealing dequeues.
\end{abstract}

\noindent\textbf{Keywords:} Distributed Algorithm, Byzantine Fault Tolerance, Conflict-free Replicated Data Types, Mazurkiewicz traces

\newcommand{\nameOfObjects}{Process-Commutative Objects\xspace}
\newcommand{\nameOfObject}{Process-Commutative Object\xspace}
\newcommand{\nameOfObjectHyph}{Process-Commutative-Object\xspace}
\newcommand{\nameOfObjectAbbr}{PCO\xspace}
\newcommand{\nameOfObjectAbbrs}{PCOs\xspace}


\section{Introduction\protect\footnote{A preliminary version of this work appeared in \cite{frey:hal-03346756}. This earlier work only covered crash faults, and contained no examples. This version generalizes the approach to Byzantine faults, presents in detail the consistency conditions offered by the approach, and applies it to four concrete examples.}}

Blockchains and, more generally, Distributed Ledger Technologies (DLT) have brought a fresh and growing interest to Byzantine Fault Tolerance (BFT), a field initiated more than 40 years ago~\cite{LSP82,PLS80}. BFT algorithms allow most modern Blockchains and DLTs to enforce a total order (TO) on their operations, even in the face of malicious participants. Unfortunately, total ordering requires solving Byzantine-Tolerant Agreement, a vexing problem that has long been known to be impossible to solve deterministically in asynchronous systems~\cite{DBLP:journals/jacm/FischerLP85}.

As a result, existing DLTs typically exhibit high execution costs and must strike a compromise regarding either termination or their guarantees~\cite{gilbert2012perspectives}---for instance, by only providing probabilistic correctness guarantees (Bitcoin), or by employing costly (randomized) synchronization mechanisms~\cite{B83,rabin1983randomized, DBLP:journals/jacm/MostefaouiMR15,rabin1983randomized}.

Total ordering and Byzantine Agreement are, however, less central to Blockchains and DLTs than first assumed. Recent works have shown that cryptocurrency objects, Blockchain's first and perhaps most emblematic application, do not require a total order and can be implemented using weaker communication abstractions, such as
FIFO reliable broadcast~\cite{auvolat:eatcs,DBLP:journals/dc/GuerraouiKMPS22,DBLP:conf/aft/BaudetDS20}\footnote{It is shown in~\cite{auvolat:eatcs} that, 
from a computability point of view, in asynchronous message-passing systems, money transfer is a weaker problem that the construction of a read/write register.}. This realization has far-reaching consequences, both practically and theoretically. In practice, it suggests that trustless money transfer applications can be implemented much more efficiently than initial implementations based on full-fledged BFT TO mechanisms (whether based on Proof-of-Work~\cite{DBLP:conf/eurocrypt/GarayKL15}, BFT consensus~\cite{DBLP:conf/sp/CrainNG21,DBLP:conf/blockchain2/CaprettoCARS22}, or Proof-of-Stake~\cite{gilad2017algorand}). From a theoretical viewpoint, it raises the question of characterizing the class of BFT distributed objects that can be directly implemented using a plain FIFO reliable broadcast. Delineating the computational power of reliable (FIFO) broadcast has in turn important ramifications, as reliable broadcast is weaker (computationally) than atomic registers\footnote{When faced with crashes, atomic registers require at least a majority of correct processes, while reliable broadcast imposes no such constraint.}.

Using weak synchronization for distributed objects is not new. It has been extensively explored in the context of \emph{Conflict-free Replicated Data Types} (CRDTs for short)~\cite{DBLP:conf/cscw/OsterUMI06,DBLP:conf/sss/ShapiroPBZ11,shapiro:inria-00555588,auvolat:hal-02303490,DBLP:conf/icde/EnesAB019,DBLP:conf/doceng/NedelecMMD13,weiss2009logoot,DBLP:journals/corr/abs-2305-16976}, a broad family of weakly synchronized distributed data structures. CRDTs eschew the need for advanced synchronization through systematic reconciliation. This pragmatic approach makes CDRTs highly scalable, robust to transient disconnections and crashes, and particularly attractive when implementing large-scale decentralized services. 

Over the last few years, several designs for Byzantine-tolerant CRDTs have been proposed that combine the strengths of traditional CRDTs with the robustness of BFT algorithms~\cite{DBLP:conf/eurosys/Kleppmann22,DBLP:conf/sicherheit/JacobBH22,DBLP:conf/fab/CholviAGNRR21,DBLP:conf/IEEEscc/Chai014}. Unfortunately, because CRDTs are not designed to block conflicting updates, these approaches cannot prevent duplicitous actions by malicious participants, which violate an application's invariant when combined. Instead, they typically remediate such attacks by canceling all or some of the incompatible operations when the problem is (eventually) detected. In particular, existing BFT CRDTs cannot prevent double spending, a problematic situation for cryptocurrency applications.

This paper is a step to chart the territory that lies between best-effort BFT CRDTs and totally ordered distributed ledgers. Using Mazurkiewicz traces, it introduces a novel class of distributed objects termed \emph{\nameOfObjects} (\nameOfObjectAbbr for short). \nameOfObjectAbbrs provide both \emph{Strong Eventual Consistency} (SEC) and \emph{Pipeline Consistency} (PC), while allowing conflicting operations under certain conditions. The paper then proposes a generic algorithm that can implement any \nameOfObjectAbbr specified by an appropriate trace language. This generic algorithm works both in a crash and Byzantine fault model depending solely on the properties of its underlying reliable broadcast. Finally the paper presents four concrete examples of \nameOfObjectAbbrs, ranging from fundamental building blocks (multisets and Petri Nets) to more advanced mechanisms (money transfer objects and work stealing deques).

\paragraph*{Roadmap}
\cref{sec:probl-stat-dlts} first presents some background on Distributed Ledgers, CRDTs, Byzantine Fault Tolerance (BFT), and recent efforts to design BFT CRDTs. \cref{sec:backgr-syst-model} moves on to define our system model, and presents the notion of \emph{trace monoid}, which is the algebraic structure that underpins the construction developed in this paper. \cref{sec:nameofobjects}  formally specifies a novel type of CRDTs that we have termed \emph{\nameOfObjects}, both in crash-prone and in Byzantine asynchronous distributed systems. Building on these definitions, \cref{sec:gener-name-algor,sec:proofs} present and prove a generic algorithm that implements any \nameOfObject.
\cref{sec:examples} discusses four practical examples of \nameOfObjects that can be implemented by this algorithm: (i) \emph{Multi-sets with Deletion Rights}, (ii) \emph{Petri Nets with Transition Rights}, (iii) \emph{Money Transfer}, and (iv) \emph{Work and Stealing Deques}.
\cref{sec:conclusion} concludes the paper.


\section{Problem Statement: DLTs, CRDTs, and Byzantine Tolerance}%
\label{sec:probl-stat-dlts}
\subsection{Blockchains and Distributed Ledgers}

Blockchains and Distributed Ledger Technologies (DLTs) have had a profound impact on both academia and industry over the last decade and have led to the rise of several highly-visible systems~\cite{nakamoto_bitcoin_2008,Ethereum2024,Riple2023,Hyperledger2023}.
These large-scale systems have found potential application in many fields beyond cryptocurrency, including self-sovereign identity (SSI) management~\cite{naik2021sovrin,MathieuSSI/DISC2023}, supply-chain networks~\cite{SupplyChainBlockchains2023}, health care applications~\cite{10.1093/jamia/ocx068}, and humanitarian relief~\cite{WFP:BB:2023}.

Distributed ledgers enforce a total order on recorded operations in an adversarial environment, i.e. an environment in which not all participants can be trusted (compromised participants are called \emph{Byzantine}). Depending on the type of ledger, participants may further join or leave freely (\emph{permissionless} systems), and/or network delays may be arbitrary (\emph{asynchronous} systems).
Despite these unfavorable circumstances, blockchains-based DLTs have succeeded in reconciling robustness, openness, and scalability, rendering them highly attractive for fields in which traceability and auditability are of prime concern.
Unfortunately, enforcing a total order in a failure-prone distributed system requires strong assumptions~\cite{DBLP:journals/jacm/FischerLP85,gilbert2012perspectives}, and is generally very costly in comparison to equivalent systems~\cite{DEVRIES2023100633}. As a result, most existing DLT systems assume some partial synchrony (e.g. the existence of some \emph{Global Stabilization Time},  GST for short)~\cite{DBLP:journals/jacm/DworkLS88} to guarantee the progress of their operations, and/or only offer probabilistic guarantees (e.g. allowing for \emph{forks} with low probability, in which recently recorded operations are canceled).

Given their inherent costs, many works have sought to improve the performance of Blockchains and DLTs at many levels. Such attempts include optimizing how transactions issued by clients are received by servers~\cite{DBLP:conf/icde/GaiNB0W23,DBLP:conf/eurosys/DanezisKSS22},  introducing some partial concurrency to order sets of operations rather than individual operations~\cite{DBLP:conf/blockchain2/CaprettoCARS22}, or making decisions in parallel on sets of transactions~\cite{DBLP:conf/sp/CrainNG21}, just to name a few recent proposals.

The above works have in common that they maintain the requirement for a global total ordering (possibly between sets of operations) that characterizes Blockchain and DLTs. Total ordering is extremely useful as it allows for linearizable distributed implementations of any sequential object~\cite{DBLP:journals/toplas/HerlihyW90,DBLP:journals/toplas/Herlihy91}, thus delivering one of the strongest consistency criteria proposed for distributed data-structures.

Many applications, however, including cryptocurrencies~\cite{auvolat:eatcs,DBLP:conf/aft/BaudetDS20,DBLP:conf/dsn/CollinsGKKMPPST20,DBLP:journals/dc/GuerraouiKMPS22}, do not require total ordering in practice. A parallel strand of works has therefore sought to alleviate the scalability and cost issues of DLTs and Blockchain by enforcing weaker ordering constrains on their operations. These work builds upon a long history of weak consistency conditions~\cite{DBLP:journals/dc/AhamadNBKH95,DBLP:conf/sosp/TerryTPDSH95,DBLP:conf/netys/FriedmanRT15,DBLP:conf/ipps/PerrinMJ15,DBLP:conf/ppopp/PerrinMJ16}, and have in particular sought to adapt a family of weakly consistent distribued data structures known as \emph{Conflict-free replicated data types} (CRDTs for short)~\cite{DBLP:conf/cscw/OsterUMI06,DBLP:conf/sss/ShapiroPBZ11,DBLP:conf/icde/EnesAB019} to the context of trustless systems initially targeted by Blockchains and DLTs~\cite{DBLP:conf/sicherheit/JacobBH22,DBLP:conf/eurosys/Kleppmann22,DBLP:conf/fab/CholviAGNRR21,DBLP:conf/IEEEscc/Chai014,OrderlessChain:2023}.

\subsection{Conflict-free replicated data types (CRDTs)}

\emph{Conflict-free replicated data types} (CRDTs)~\cite{DBLP:conf/cscw/OsterUMI06,DBLP:conf/sss/ShapiroPBZ11,DBLP:conf/icde/EnesAB019,weiss2009logoot,DBLP:conf/popl/GotsmanYFNS16} were initially proposed for crash-prone large-scale peer-to-peer and geo-replicated systems.
A CRDT provides update and read operations. Its state is replicated on all participants, and kept synchronized using a best-effort reconciliation strategy. Crucially, the operations of a CRDTs are assumed to commute, so that the order in which they are applied onto a replica neither impacts the resulting state, nor the legality of possible future operations.
Formally, if $\op_1,\op_2\in\mathcal{O}$ are two update operations supported by a CRDT, and $S\in\mathcal{S}$ is one of the CRDT's reachable states, then
\begin{equation}\label{eq:CRDT:commutativity}
  \valid(\op_1,S) \wedge \valid(\op_2,S) \text{ implies }
  \left\{\begin{array}{@{}l}
    \valid(\op_1,S\cat \op_2) \wedge\\
    \valid(\op_2,S\cat \op_1) \wedge\\
    (S\cat \op_1) \cat \op_2 = (S\cat \op_2) \cat \op_1,
  \end{array}\right.
\end{equation}
where $\valid(\op_x,S)$ indicates that operation $\op_x$ is legal\footnote{Shapiro, Preguiça, Baquero, and Zawirski~\cite{DBLP:conf/sss/ShapiroPBZ11,shapiro:inria-00555588} use the notion of \emph{precondition} rather than validity; the two are, however, closely related. In particular, when an operation $\op_x$ that has been locally applied is disseminated to replicas (in the \emph{downstream phase} using the vocabulary of \cite{DBLP:conf/sss/ShapiroPBZ11,shapiro:inria-00555588}), saying that $\op_x$'s \emph{downstream precondition} is enabled in the state $S$ of a replica means that $\op_x$ is valid in $S$.} when the CDRT is in state $S$, and $S\cat \op_x$ represents the state resulting from the application of operation $\op_x$ on state $S$.

CRDTs can typically be implemented using two main designs:
\begin{itemize}
\item In op-based implementations~\cite{DBLP:conf/cscw/OsterUMI06,DBLP:conf/sosp/PetersenSTTD97,DBLP:conf/sosp/JosephdTGK95}, update operations are broadcast to all replicas which apply them as soon as they become legal on their local state. Op-based implementations requires an underlying reliable broadcast service that guarantees that all updates are eventually delivered to all replicas.
\item In state-based implementations~\cite{DBLP:conf/sosp/DeCandiaHJKLPSVV07}, the set of reachable states $\mathcal{S}$ must be a \emph{join semilattice}, i.e. a partial order $(\mathcal{S},<_{\mathcal{S}})$ equipped with a least-upper-bound operator $\sqcup_{\mathcal{S}}$, such that for any pair of states $S_1,S_2\in\mathcal{S}$, $S_{\sqcup}=S_1\sqcup_{\mathcal{S}} S_2$ is the smallest state of $S$ according to $<_{\mathcal{S}}$ that is larger than both $S_1$ and $S_2$. In this design replicas apply updates locally and periodically propagate their state to other replicas (possibly directly, but most often using some stochastic exchanges known as gossiping~\cite{demers_epidemic_1987}). When a replica $p_i$ receives the state $S_j$ of another replica $p_j$, $p_i$ simply updates its own local state $S_i$ using the least-upper-bound operator $\sqcup_{\mathcal{S}}$
  \begin{equation*}
    S_i \gets S_i \sqcup_{\mathcal{S}} S_j.
  \end{equation*}
  A canonical example of a state-based CRDTs is a \emph{grow-only set} over a universe $U$ of elements that is implemented using $\mathcal{S}=\mathcal{P}(U)$ (the powerset of $U$) as set of states, $\subseteq$ as partial order $\mathcal{S}$, and $\cup$ as least-upper-bound operator.
\end{itemize}

Both designs ensure that a CRDT is eventually consistent, i.e., that if replicas stop issuing updates, then all replicas eventually converge towards the same state~\cite{DBLP:journals/csur/ViottiV16,DBLP:conf/podc/DuboisGKPS15,DBLP:conf/sosp/TerryTPDSH95}.
State-based CRDTs are typically easier to reason about than op-based approaches, and only need a dissemination mechanism with very weak properties. Sending whole states over the network can, however, incur large overheads, a problem that is often addressed through delta-based implementations, where only part of the state is exchanged between peers~\cite{auvolat:hal-02303490,DBLP:conf/icde/EnesAB019}. 

CRDT algorithms have been proposed for a broad range of distributed objects ranging from sets (including Grow-Only Sets, 2-Phase sets, Observed-Removed Sets~\cite{shapiro:inria-00555588}) to collaborative editors~\cite{DBLP:conf/doceng/NedelecMMD13,weiss2009logoot,DBLP:conf/cscw/OsterUMI06} through money-transfer distributed ledgers~\cite{DBLP:journals/corr/abs-2305-16976}.
In particular, the work presented in~\cite{DBLP:journals/corr/abs-2305-16976} proposes an algorithm that implements a money transfer distributed ledger as a state-based CRDTs by relying on grow-only counters (GOCs). The proposed approach tolerates crashes, and allows negative accounts (which are usually excluded in cryptocurrency applications). It hinges on a data structure that captures a sets of accounts, whose (overall) state space forms a join semilattice. The data structure keeps track of the flow of money between accounts by dissociating the sending of token from their reception.

\subsection{Byzantine Fault-Tolerant CRDTs}

The flexibility and low cost of CRDTs have led to attempts to circumvent the constraints imposed by most Blockchains and DLTs by adding Byzantine Fault-Tolerance to CRDTs~\cite{DBLP:conf/eurosys/Kleppmann22,DBLP:conf/sicherheit/JacobBH22,DBLP:conf/fab/CholviAGNRR21,DBLP:conf/IEEEscc/Chai014,OrderlessChain:2023}.
For instance, in~\cite{DBLP:conf/fab/CholviAGNRR21} the authors formally specify \emph{Byzantine Grow-only Set Object} (BDSO), and propose an algorithm that implement BDSO based on the definition of Grow-Only Sets (G-Sets) initially introduced for crash-only environments~\cite{DBLP:conf/sss/ShapiroPBZ11}. The proposed algorithm assumes a client/server setting in which $n$ servers are replicated, any client may be Byzantine, and up to $t<n/3$ servers may be Byzantine. It relies on an underlying Byzantine Reliable Broadcast~\cite{raynal-book18,LSP82} and on signatures to implement a grow-only sets that is eventually consistent from the point of view of correct clients.

Adopting a peer-to-peer setting, the approach presented in~\cite{DBLP:conf/eurosys/Kleppmann22} provides a generic technique to render op-based CRDTs tolerant to an arbitrary number of Byzantine participants. The proposed algorithms assumes that an update's validity only depends on its causal past (a common assumption for CRDTs), and that all updates commute. It encodes updates in a Direct Acyclic Hash Graph that captures causality relations between updates, and guarantees that all correct replicas eventually converge towards the same state.

\subsection{Problem statement}
\label{sec:problem-statement}
\newcommand{\munit}{\textsf{mu}\xspace}%

Existing Byzantine Fault-Tolerant CRDTs require that update operations commute, and crucially that their legality is not impacted by the order in which update operations are applied. This excludes application such as cryptocurrency in which a token (or money in general) may not be used twice (thus forbidding double-spending attacks).

As a simple example, assume Alice owns 10 monetary units (\munit). Alice may perform a transfer of 10\,\munit to Bob. She may also transfer 10\,\munit to Clara. Alice may not, however, perform both transfers. Once she has transferred 10\,\munit to Clara, the transfer to Bob of the same amount becomes illegal, as negative accounts are forbidden.
This type of constrain on the legality of operations breaks the core assumption of commutativity that underpins traditional CRDTs as captured by~\cref{eq:CRDT:commutativity}. There is however hope. The remainder of this work presents a systematic approach to model and enforce constrains on forbidden states despite Byzantine participants without resorting to a total order to resolve conflicts.

\section{Background: System Model and the Trace Monoid}
\label{sec:backgr-syst-model}
	\newcommand{\CAMP}{{\mathit{CAMP}_{n,t}}}
	\newcommand{\BAMP}{{\mathit{BAMP}_{n,t}}}

Our approach assumes a message-passing asynchronous model with a fixed number of participants, and considers two fault models (processes crashes and Byzantine processes). In the following we formally define this computational model. We then provide some basic definitions on Mazurkiewicz Traces and trace languages, which we use to specify \nameOfObjects in \cref{sec:nameofobjects}.
  
  
\subsection{Message-passing computation in the presence of Byzantine advsersaries}
\label{sec:mess-pass-comp}
We consider a distributed system that is composed of $n$ asynchronous processes $\Pi=\{p_1$, ..., $p_n\}$. `Asynchronous processes' means that each process proceeds at its own speed, which can vary with time and remains always unknown to other processes.

Each pair of processes is connected by an asynchronous reliable bi-directional channel. `Asynchronous channel' indicates that the transfer delay of each message, taken individually, is arbitrary but finite, `reliable' means that there is neither message corruption, creation, nor message losses. Channels are further assumed to be authenticated, i.e. when a process $p_i$ receives a message $m$ from $p_j$ over the channel $c_{i,j}$, $p_i$ knows $m$ originated from $p_j$, and this information cannot be compromised.

This work considers two types of process failures, process crashes and Byzantine processes.
\begin{itemize}

\item A \emph{process crash} is a premature and definitive stop of a process while it is executing its algorithm. We will use the notation $\CAMP[\emptyset]$ (where CAMP stands for \emph{Crash Asynchronous Message Passing}) to denote the above asynchronous distributed computing model in which up to $t<n$ processes may crash, and the notation $\CAMP[t<a]$ when the maximum number of processes allowed to crash is restricted to be strictly smaller than $a$.
  
\item  A \emph{Byzantine process} is a process that behaves arbitrarly, i.e., a process that does not follow the code it is assumed to execute, but remains restricted to sending and receiving messages. A Byzantine process in particular cannot prevent communication between two correct processes, or spawn new processes. Several Byzantine processes may however collude or control message delays to their advantage. In the following, the notation $\BAMP[\emptyset]$ (where BAMP stands for \emph{Byzantine Asynchronous Message Passing}) denotes the above asynchronous distributed computing model in which up to $t<n$ processes may be Byzantine, while $\BAMP[t<a]$ is used to denote the same model when the maximum number of Byzantine processes is restricted to be strictly smaller than $a$.

\end{itemize}

\subsection{Mazurkiewicz Traces}
\label{sec:mazurkiewicz-traces}

The family of distributed objects we introduce in \cref{sec:nameofobjects} hinges on the notion of \emph{trace language} which is itself derived from that of \emph{trace monoid}.
A trace monoid~\cite{10.5555/25542.25553,DBLP:conf/ershov/Zielonka89,DBLP:books/ws/95/DR1995,DBLP:reference/parallel/DiekertM11} (or free partially commutative monoid, also known as Mazurkiewicz Traces) is an algebraic structure that generalizes the notion of words (i.e., finite sequences) over an alphabet $\Sigma$ and naturally captures parallelism and conflicts between operations (represented by the letters of the alphabet).

A trace monoid over an alphabet $\Sigma$ is defined by a symmetric independency relation $I \subseteq \Sigma \times \Sigma$ between the letters (or operations) of $\Sigma$. $(a,b)\in I$ means that the operations $a$ and $b$ commute, so that the effect of $ab$ and $ba$
are equivalent. Two (finite) words $u,v \in \Sigma^{\ast}$ are said to be equivalent under $I$, noted $u \sim_I v$ iff one can transform $u$ into $v$ (and reciprocally) by exchanging adjacent operations that are independent within $u$.

The relation $\sim_I$ is an equivalence relation over $\Sigma^{\ast}$, and a (finite) trace $t$ is simply an equivalence class of $\sim_I$ within $\Sigma^{\ast}$. $\sim_I$ is a congruence for the concatenation operator (noted $\cat$, but generally omitted), i.e if $x \sim_I y$ and $u \sim_I v$, then $xu \sim_I yv$. As a result, the concatenation over words translates to the set of traces, with $[u]_I[v]_I = [uv]_I$, where $u, v\in \Sigma^{\ast}$ are words over $\Sigma$, and $[u]_I$ is the trace represented by $u$ ($u$'s equivalence class under $\sim_I$). The resulting structure $\left(\faktor{\Sigma^{\ast}}{\sim_I},\cat\right)$ is the free partially commutative monoid defined by $\Sigma$ and $I$, noted
$\mathbb{M}(\Sigma,I)$, or $\mathbb{M}(\Sigma)$ for short. As for words, a subset of $\mathbb{M}(\Sigma,I)$ is called a \emph{trace language}.

In the following, unless ambiguous, we will usually note a trace using one of its representative elements (i.e. $ab$ instead of $[ab]_I$).

\section{\nameOfObjects}
\label{sec:nameofobjects}

\nameOfObjects (\nameOfObjectAbbr for short) offer convergence properties that go beyond those of CRDTs, with the additional ability to capture and enforce constrains on forbidden states despite failures.
This section formally defines this new family of distributed objects using execution histories (a common approach to describe distributed objects) and trace languages, which we have just discussed. We progress incrementally, first introducing the operations supported by a \nameOfObjectAbbr (\cref{sec:operations}), before moving on to a sequential specification (\cref{sec:sequ-spec}), and finally providing a distributed specification of a \nameOfObjectAbbr in both the crash and Byzantine failure models (\cref{sec:few-remarks-diff,sec:distr-name}).

The interest of \nameOfObjects comes both from the consistency guarantees they offer and the ease with which they can be implemented in adversarial environments. \cref{sec:gener-name-algor} introduces a generic algorithm that implements any \nameOfObjectAbbr. The same algorithm works both in crash and Byzantine failure models, depending only on the type of reliable broadcast that it uses underneath.

\subsection{Operations}
\label{sec:operations}

A \nameOfObject is a distributed data structure that accepts two types of operations, \emph{updates} and \emph{queries}, while obeying a small number of rules on the effect and interaction of these operations.
  
\begin{itemize}
\item \emph{Updates} modify the data structure. Updates might return a value (which reflects the data structure's state). If they do not, they are said to be \emph{silent}. The set of update operations is noted $\Sigma$.
\item \emph{Queries} do not modify the data structure, but return a value, that deterministically depends on the data structure's state. The set of query operations is noted $Q$.
\end{itemize}

Any process may execute any query, but some update operations are
only allowed to some processes. More specifically, the set of updates
$\Sigma$ is made of $n+1$ subsets, $\Sigma= C \cup \bigcup_{p_i\in \Pi}O_i$, where
\begin{itemize}
\item $C$ is the set of \emph{common updates}, which are accessible to all processes, and commute with any other operations;
\item $O_i$ are updates that are exclusive to process $p_i$, and do not commute with each other.
\end{itemize}

The sets $C$, and $O_i$ induce an independency relation $I$ over $\Sigma$:
\begin{equation}\label{eq:indep:rek:definition}
  I =\Sigma\times\Sigma \setminus \bigcup_{p_i\in \Pi}(O_i \times O_i).
\end{equation}

Because the pairs of operations included in $I$ commute, the effects of a sequence of operations on a  \nameOfObjectAbbr can be summarized through a trace of the monoid $\mathbb{M}(\Sigma,I)$ (noted $\mathbb{M}(\Sigma)$ for short in the following), and the state space of a \nameOfObjectAbbr can be equated with the trace monoid $\mathbb{M}(\Sigma)$.\footnote{In practice, using $\mathbb{M}(\Sigma)$ as a state space assumes that a \nameOfObjectAbbr has a well-defined initial value, associated with the empty trace $\epsilon$. (We will generally conflate the empty trace and the empty sequence, and note both of them $\epsilon$.) Note that actual implementations of a \nameOfObjectAbbr should generally not use $\mathbb{M}(\Sigma)$ to represent the \nameOfObjectAbbr's state space, but should opt for a compact application-dependent representation. Referring to $\mathbb{M}(\Sigma)$ is however useful at an abstract level, as it provides reasoning tools that are both generic and expressive.}

Given some update operation $\update\in\Sigma$, we note $\outputFunc(\update,v)$ the value returned by $\update$ when it is applied to the state captured by the trace $v\in \mathbb{M}(\Sigma)$:
  \begin{equation*}
  \begin{array}{cccc}
    \outputFunc:&\Sigma\times\mathbb{M}(\Sigma)&\to&\mathbb{V},\\
        &(\update,v)&\mapsto& \outputFunc(\update,v),
  \end{array}
  \end{equation*}
  where $\mathbb{V}$ is a set of possible output values.
  When $\update$ is silent, $\outputFunc(\update,v)$ should return some pre-determined constant value (e.g. $\bot$) for all $v$.

\subsection{Sequential Specification}
\label{sec:sequ-spec}

\subsubsection{Miscellaneous notations}

In the following we will note $[1..n]$ the set of natural numbers from $1$ to $n$ (included); $(x_j)_{j\in[1..n]}$ the sequence $(x_1, ..., x_n)$; and $\{x_j\}_{j\in[1..n]}$ the set $\{x_1, ..., x_n\}$. When the context is clear, for brevity, we will abbreviate $(x_j)_{j\in[1..n]}$ in $(x_j)_j$ and $\{x_j\}_{j\in[1..n]}$ in $\{x_j\}_j$.

\subsubsection{\nameOfObjectAbbr specification}

  A \nameOfObject is specified by a tuple $\big(C,(O_j)_{j\in[1..n]},L,Q\big)$ where
\begin{enumerate}[i)]
\item
 $C$ and $(O_j)_j$ are the sets of common and process-specific updates
  introduced just before,
\item $L$ is a language of finite traces over the free partially commutative monoid $\mathbb{M}(\Sigma,I)$ defined over the object's update operations $\Sigma = C \cup \bigcup_{p_i\in \Pi}O_i$ with the independence relation $I\subseteq \Sigma\times\Sigma$ defined
  earlier,
\item and $Q=\{q_k\}_{k\in[1..|Q|]}$ is a set of query functions $q_k: L \rightarrow
 \mathbb{V}$, where $\mathbb{V}$ is a set of application-dependent output values.
\end{enumerate}  
  $L$ must further respect the following three properties:
	\begin{itemize}
  \item {\bf Initial emptiness:} the empty trace is part of $L$: $[\epsilon]_I\in L$.
	\item {\bf $C^{\ast}$-Closure:} $L$ is closed by concatenation with traces of common updates: $LC^{\ast} \subseteq L$.
  \item {\bf $I$-diamond Closure:}
    Given $v \in L$, an update $\op_i \in O_i$ exclusive to some process $p_i$, and some trace of updates $z \in \mathbb{M}(\Sigma \setminus O_i)$ that are independent of $\op_i$, if
    $\{v\cat \op_i, v\cat z\}\subseteq L$ then $v\cat z \cat op_i \in L$.
\end{itemize}

  \emph{Initial emptiness} states that the initial state of a \nameOfObjectAbbr must be valid. \emph{$C^{\ast}$-Closure} indicates that common updates do not change the legality of future updates. (Updates in $C$ can be seen as following the validity rule of classical CRDTs as captured in \cref{eq:CRDT:commutativity}.) Finally, \emph{$I$-diamond Closure} means that an update $\op_i$ that is exclusive to $p_i$ does not change the legality of future updates by other processes (which include common updates and updates exclusive to processes other than $p_i$). The application of $\op_i$ may, however, impact the legality of future updates exclusive to $p_i$. For instance, returning to the example of \cref{sec:problem-statement}, only Alice is allowed to transfer the 10\,\munit she owns, but after transferring 10\,\munit to Clara, Alice may no longer be able to transfer 10\,\munit to Bob.
 \subsubsection{\nameOfObjectAbbr-legal sequential execution}
 \label{sec:name-legal-sequ}
 
 As is typical when specifying distributed objects and consistency conditions~\cite{DBLP:journals/dc/AhamadNBKH95,DBLP:journals/toplas/HerlihyW90,DBLP:conf/ipps/PerrinMJ15,DBLP:conf/ppopp/PerrinMJ16,DBLP:conf/netys/FriedmanRT15}, we first specify the sequential execution of a \nameOfObject. The following two sections then build on this sequential specification to define the acceptable behavior of \nameOfObjectAbbrs in a distributed system in the presence of either crashes (\cref{sec:PCO-CAMP}) or Byzantine processes (\cref{sec:PCO-BAMP}).
 
 Formally, a sequential execution $S$ of a \nameOfObject described by
the tuple $\big(C,(O_j)_j,L,Q\big)$ is a sequence of \emph{invocations} $S=(e_k)_{k\in[1..|S|]}$,
 where each invocation is labeled by an operation $\op$ from
 $\Sigma\cup Q$, a return value $\val$ from
 $\mathbb{V}\cup\{\unknown\}$ ($\unknown\not\in\mathbb{V}$ represents
 a \emph{masked} value, which will serve to specify distributed
 \nameOfObjectAbbr executions; values that are not masked are said to
 be \emph{visible}), and the process $p$ that invokes the
 operation. We note $e_k=(\op,\val,p)$.\footnote{This is equivalent to
 defining sequential executions as words over the alphabet
 $(\Sigma\cup
 Q)\times(\mathbb{V}\cup\{\unknown\})\times\Pi$.%
 }

 $S$ is \emph{\nameOfObjectHyph-legal} (\nameOfObjectAbbr-legal for short) w.r.t. $\big(C,(O_j)_j,L,Q\big)$ if the following holds:
 \begin{itemize}
 \item {\bf Process Authorization.}
   Each operation is invoked by a process authorized to execute it: 
   $\forall(\op,\val,p_i)\in S$, $\op\in C\cup O_i \cup Q$.
   
 \item {\bf Update Legality.} Each prefix of updates of $S$ represents a legal trace:
   \begin{equation*}
     \big\{[u]_I \bigm| \exists v\in \Sigma^{\ast} : u\cat v=\proj{S}{\Sigma} \big\}\subseteq L,
   \end{equation*}
   where $\proj{S}{\Sigma}$ is the sequence of update operations in $S$;
 \item {\bf Output and Query Validity.} Each query reads from the sequence of updates up to its invocation: if $e_k= (q,\val,p)$,  with $q\in Q$, then $\val = q\big([\proj{S_{<k}}{\Sigma}]_I\big)$, where $S_{<k}=(e_\ell)_{\ell\in[1..k-1]}$ is the sequence of invocations that precede $e_k$.

Similarly, each update associated with a visible value (i.e. $\val\neq\unknown$) returns the output obtained from the sequence of updates up to the update's invocation: if $e_k=
(\update,\val,p)$ with $\update\in \Sigma$ and $\val\neq\unknown$, then $\val =\outputFunc\big(\update,[\proj{S_{<k}}{\Sigma}]_I\big)$, where $S_{<k}$ is defined as above.

\end{itemize}

 \emph{Process Authorization} enforces that the operations in $O_i$ can only be invoked by $p_i$. \emph{Update Legality} ensures that the sequence $S$ is indeed reachable from the initial state (captured by the empty sequence $\epsilon$). For instance for the sequence $S=abcd$ to respect update legality, then $\{ [a]_I, [ab]_I, [abc]_I, [abcd]_I \}$ must be included in $L$. If it is not, then one of the intermediate states encoded by $S$ is illegal according to the trace language $L$, and $S$ cannot be reached. Finally, \emph{Output and Query Validity} means that any returned value takes into account all updates in the sequence up to the point where this return value is generated.

\subsection{A few remarks on the difference between a trace and a word language}
\label{sec:few-remarks-diff}

The specification presented in the previous section follows closely the approach used to define sequential objects using word languages (i.e. subsets of $\Sigma^{\ast}$). As with words, a trace $u\in L$ can be understood as a labeled order of operations that contains enough information to determine the \emph{state} of a \nameOfObjectAbbr object $o$. More specifically, $u$ captures enough information to determine the output of any query function, and together with $L$, $u$ determines which subsequent operations are legal (any operation $\op$ such that $u\cat \op \in L$).

The use of traces rather than words introduces however some subtle yet significant changes. In a word-based specification, the sequence of operations that produces a word $s$ is unique. As a result, if $s$ is legal, any prefix of $s$ will typically be legal as well (otherwise $s$ cannot be reached). This is not the case with traces. Because the order between operations that is encoded in a trace is partial rather than total, the sequence of successive operations that leads to a trace $u$ is not unique, and the fact that a trace $u$ is legal does not imply that all its prefixes necessarily are.

\newcommand{\tab}{t_{AB}}
\newcommand{\tba}{t_{BA}}

As an illustration, consider a simple token ring object with two processes $p_A$ (for Alice) and $p_B$ (for Bob), and two update operations $\tab$ (the token goes from Alice to B) and $\tba$ (the token goes from Bob to Alice). A token ring in which the token starts with Alice can be encoded by the trace language $L = \big\{u\in \mathbb{M}(\Sigma,I) \bigm| 0 \leq \multip{u}(\tab)-\multip{u}(\tba)\leq 1 \big\}$ defined on the alphabet $\Sigma=\{\tab,\tba\}$, with the independence relation $\tab\sim_I \tba$, and where $\multip{u}(\op)$ indicates the number of times the operation $\op$ appears in the trace $u$.

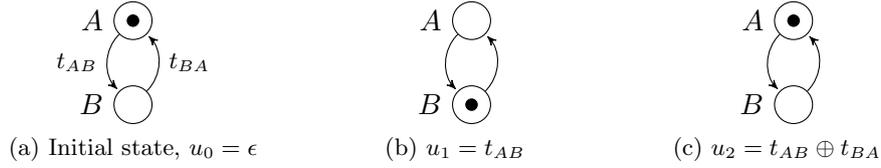
\begin{figure}[h]\centering
  \newcommand{\subfigwidth}{0.25\linewidth}
  \subfloat[Initial state, $u_0=\epsilon$]{
    \parbox{\subfigwidth}{\centering
      \begin{tikzpicture}[node distance=0.6cm and 0.7cm,
          >=stealth',bend angle=45,auto]
        \node [place,tokens=1,label=left :$A$  ] (a)  {} ;
        \node [place,         label=left :$B$  ] (b)  [below=of a] {};
        \path[->]  
        (a) edge[post,bend right] node[left , font=\footnotesize] {$t_{AB}$} (b)
        (b) edge[post,bend right] node[right, font=\footnotesize] {$t_{BA}$} (a);
      \end{tikzpicture}
    }
  }
  \subfloat[$u_1=t_{AB}$]{
    \parbox{\subfigwidth}{\centering
      \begin{tikzpicture}[node distance=0.6cm and 0.7cm,
          >=stealth',bend angle=45,auto]
        \node [place,        , label=left :$A$  ] (a)  {} ;
        \node [place,tokens=1, label=left :$B$  ] (b)  [below=of a] {};
        \path[->]  
        (a) edge[post,bend right]  (b)
        (b) edge[post,bend right]  (a);
      \end{tikzpicture}
    }
  }
  \subfloat[$u_2=t_{AB}\cat t_{BA}$]{
    \parbox{\subfigwidth}{\centering
      \begin{tikzpicture}[node distance=0.6cm and 0.7cm,
          >=stealth',bend angle=45,auto]
        \node [place,tokens=1, label=left :$A$  ] (a)  {} ;
        \node [place,        , label=left :$B$  ] (b)  [below=of a] {};
        \path[->]  
        (a) edge[post,bend right]  (b)
        (b) edge[post,bend right]  (a);
      \end{tikzpicture}
    }
  }
  \caption{An execution of a toy token ring specified over the alphabet $\Sigma=\{t_{AB},t_{BA}\}$ by the trace language $L = \big\{u\in \mathbb{M}(\Sigma,I) \bigm| 0 \leq \multip{u}(\tab)-\multip{u}(\tba)\leq 1 \big\}$. }\label{fig:example:token:ring}
\end{figure}

\cref{fig:example:token:ring} shows an execution of this toy example with two token exchanges. \cref{eq:u5} represents the trace $u_5$ obtained after five exchanges. In the equation, $u_5$ is shown as a Hasse diagram, with smaller elements on the left. $u_5$ is a legal trace of $L$ and the only legal action after $u_5$ is $\tba$. By contrast, the trace $v$ shown in \cref{eq:v} is not legal (does not belong to $L$)---as Bob cannot transfer twice the token to Alice without ever receiving it from her.

\newcommand{\predrel}{\rightarrow}
\newcommand{\negspace}{\hspace{-0.1em}}
\begin{align}\label{eq:u5}
  u_5 = &\left(\negspace\vcenter{\hbox{\begin{tikzpicture}
  \node[anchor=west] at (0,0.6) {$\tab \predrel \tab \predrel \tab$};
  \node[anchor=west] at (0,0) {$\tba \predrel \tba$};
  \end{tikzpicture}}}\negspace\right),\\
\label{eq:v}
  v = &\;\left(\negspace\vcenter{\hbox{\begin{tikzpicture}
  \node[anchor=west] at (0,0) {$\tba \predrel \tba$};
  \end{tikzpicture}}}\negspace\right).
\end{align}

This may be surprising as $v$ is a trace prefix of $u_5$ (since $u_5=v\cat t_{AB}\cat t_{AB}\cat t_{AB}$), and seems to lead to a paradox in which a legal trace can be built by appending operations to an illegal one. This paradox becomes easier to apprehend if we return to the initial ``constructionist'' definition of traces, as equivalence classes of the relation $\sim_I$ on words (\cref{sec:mazurkiewicz-traces}). Under this definition, although the word $w=\tba\tba\tab\tab\tab$ belongs to the trace $u_5$ (which is legal), it does not represent a legal sequence of operations on the token-ring object, as $w$ does not start with $\tba$ and does not alternate $\tab$ and $\tba$. The key point to note here is that the fact that the trace $u_5$ is legal (i.e., $u_5\in L$) does not imply that all representative words of $u_5$ are (only some are, typically). The use of traces however captures the indistinguishably (in terms of queries and future operations) between those words of $u_5$ that are possible under $L$. (These are the words $w\in u_5$ whose all prefixes are legal according to $L$ when considered as traces, i.e. $w\in u_5$ such that $\big\{[v]_I\;\big|\;v\text{ is a prefix of }w\big\}\subseteq L$.)

  \subsection{Distributed Histories, Perceived Histories}

  The notion of history is commonly used to model a distributed execution between processes~\cite{DBLP:journals/toplas/HerlihyW90,DBLP:conf/ipps/PerrinMJ15,DBLP:conf/ppopp/PerrinMJ16}. We further introduce the notion of \emph{perceived history} of a process $p_i$ using the projection introduced in~\cite{DBLP:conf/ppopp/PerrinMJ16}, which removes all values returned by processes other than $p_i$. (In our definition, queries are simply deleted, and the return value of updates is set to the unknown value $\unknown$.)
  
	\subsubsection{Histories}
  \label{sec:histories}
  \begin{itemize}
	\item A local execution (or history) of a process $p_i$,  denoted $E_i$, is a sequential execution $(e_k)_k=(\op_k,\val_k,p_i)_k$ in which every operation is invoked by $p_i$, and authorized to $p_i$, i.e. $\op_k \in C \cup O_i \cup Q$, and each returned value is known, i.e. $\val_k\neq\unknown$.

  If an invocation $e_{k_1}=(\op_{k_1},\val_{k_1},p_i)$ precedes an invocation $e_{k_2}=(\op_{k_2},\val_{k_2},p_i)$ (i.e. $k_1 < k_2$), we write $e_{k_1} \rightarrow_i e_{k_2}$, or for short $\op_{k_1} \rightarrow_i \op_{k_2}$. $\rightarrow_i$ is the \emph{local process order} for $p_i$.
    
	\item A history $H$ is a set of $n$ local histories, one per process, $H = (E_1,\dots, E_n)$.

	\item Given a history $H$ and a process $p_i$, we define $p_i$'s \emph{perception} of $H$ ($p_i$'s perceived history of $H$) as the history $\widehat{H}_i$ in which all query operations except those of $p_i$ have been removed and the output of update operations by processes other than $p_i$ have been masked. Formally, $\widehat{H}_i = (\widehat{E}_1, \dots, \widehat{E}_n)$ with:
	\begin{itemize} 			
    \item $\widehat{E}_i = E_i$,
    \item $\widehat{E}_j = \proj{E_j}{(\Sigma\times\{\unknown\}\times\{p_j\})}$,
	\end{itemize}
  where $\proj{E_j}{(\Sigma\times\{\unknown\}\times\{p_j\})}$ represents the sequence of invocations obtained by removing all queries from $E_j$, and by masking the output values of the remaining updates, i.e. replacing $e_k=(\update_k,\val_k,p_j)\in E_j$ by $e'_k=(\update_k,\unknown,p_j)$. We say that the output of $e_k$ is \emph{masked} in $\widehat{E}_j$. (This transformation is similar to the ``hiding'' projection introduced in~\cite{DBLP:conf/ppopp/PerrinMJ16}.)

\end{itemize}

\subsubsection{Serialization}
\begin{itemize}
\item A \emph{serialization} $S$ of a history $H$ is a sequential execution which
		contains all invocations of $H$ and respects all local process orders
		$\{\rightarrow_i\}_{i\in[1..n]}$.
\item Given a history $H$ and a process $p_i$, a \emph{local serialization} $S_i$ is a
		serialization (topological sort) of the history $\widehat{H}_i$.
\end{itemize}

\subsection{Distributed \nameOfObjects}\label{sec:name-compl}
\label{sec:distr-name}

We have now set the stage needed to specify how a \nameOfObjectAbbr should behave in a distributed setting. We provide two specifications; the first assumes a crash model ($\CAMP[\emptyset]$ using the notation of \cref{sec:mess-pass-comp}) and the other works in a Byzantine setting ($\BAMP[t < n/3]$). The two specifications differ in the way faulty processes are handled. Whereas in the crash model we can define the desirable behavior of faulty processes up to the point when they crash, in the Byzantine case the local invocations performed by a Byzantine node cannot be constrained in any way.
The different constraints on $t$ (the maximum number of faulty processes) reflect the computability bounds on reliable broadcast, which we use in \cref{sec:gener-name-algor} to implement \nameOfObjects. In a crash model, reliable broadcast can be implemented for any number of faulty processes (there is no constrain on $t$). By contrast, in a Byzantine process model, reliable broadcast requires that $t<n/3$~\cite{bracha1987asynchronous}.

	\subsubsection{\nameOfObjects in \texorpdfstring{$\CAMP[\emptyset]$}{CAMP}}
\label{sec:PCO-CAMP}

With crashes, the last update operation invoked by a crashed process is not necessarily seen by other processes (correct or crashed). To model this situation, the definition below introduces the notion of \emph{crash-corrected history}, in which the last update of crashed processes might be removed.  
The remainder of the definition insures that each process $p_i$ eventually sees all updates issued by other processes (correct or not), minus possibly the last update of crashed processes. The returned value of local updates and queries performed by a process $p_i$ must further be explainable by a serialization $S_i$ of the history perceived by $p_i$ in which the outputs of updates performed by remote processes $p_j\neq p_i$ are masked~\cite{DBLP:conf/ppopp/PerrinMJ16} (\cref{sec:histories}). In particular, these masked outputs do not need to agree with $S_i$.

More formally, we define	a \emph{crash-corrected history} associated with an execution $H$ of a \nameOfObject as the history $H' = (E'_1, \dots, E'_n)$ such that:
\begin{itemize}
\item $E'_i = E_i$ if $p_i$ is correct,
\item $E'_i$ is either equal to $E_i$, or equal to $E_i$ minus its
last element if $p_i$ is faulty (local crash-corrected history).
\end{itemize}

\begin{definition}[\nameOfObjectAbbr-Compliance in the crash failure model]\label{def:PCO:compliance:CAMP}
In the $\CAMP[\emptyset]$ model, a distributed objects implements
the \nameOfObjectAbbr specification $\big(C,(O_j)_j,L,Q\big)$ if it provides
each process $p_i$ with the operations contained in $C \cup O_i \cup Q$
and is such that, for each of its executions, represented by the
corresponding history $H$, we have:

\begin{itemize}
\item All the operations invoked by correct processes terminate.
\item There is a crash-corrected history $H'$ of $H$ such that for each process $p_i$ there is a \nameOfObjectAbbr-legal serialization $S_i$ of $\widehat{H'}_i$, where $\widehat{H'}_i$ is $p_i$'s perceived history of $H'$ defined in \cref{sec:histories}.
\end{itemize}

\end{definition}

\subsubsection{\nameOfObjects in \texorpdfstring{$\BAMP[t < n/3]$}{BAMP}}
\label{sec:PCO-BAMP}
The notion of local history $E_i$ is no longer meaningful for Byzantine processes because a Byzantine process can not only crash but can also simulate, omit, or otherwise disrupt operations, either invoked by itself or by others. In particular, there is no relation on a Byzantine process $p_b$ between the operations that an application executing on $p_b$ might invoke and the behavior that $p_b$ exhibits on the network in terms of messages. To circumvent this disconnect, we introduce the notion of \emph{mock local history} for Byzantine processes. A \emph{mock local history} of a Byzantine process $p_b$ is a local history of $p_b$ that $p_b$ could have produced if it were correct, and that \emph{explains} $p_b$'s behavior from the point of view of the system's correct processes. The system's correct processes implement a \nameOfObjects if they all behave in a manner consistent with the same set of mock local histories for the Byzantine processes.

More formally, we define	a \emph{mock history} associated with an execution $H$ of a \nameOfObject as the history $H' = (E'_1, \dots, E'_n)$ such that:
\begin{itemize}
\item $E'_i = E_i$ if $p_i$ is correct,
\item $E'_i \subseteq (O_i \cup C)^{\ast}$ if $p_i$ is Byzantine (mock
local history). \end{itemize}

\begin{definition}[Distributed \nameOfObjectAbbr in the Byzantine failure model]
A \nameOfObjectAbbr implements the \nameOfObjectAbbr specification $\big(C,(O_j)_j,L,Q\big)$ in $\BAMP[t < n/3]$ if it provides each correct process $p_i$ with the operations contained in $C \cup O_i \cup Q$ and is such that, for each of its executions, represented by the corresponding history $H$, we have:

\begin{itemize}
\item All the operations invoked by correct processes terminate.
\item There is a mock history $H'$ of $H$ such that for each correct process $p_i$ there is a \nameOfObjectAbbr-legal serialization $S_i$ of $\widehat{H'}_i$, where $\widehat{H'}_i$ is $p_i$'s perceived history of $H'$ defined in \cref{sec:histories}.
\end{itemize}
\end{definition}

\subsubsection{Discussion: Consistency of \nameOfObjects}

The above definitions ensure that non-Byzantine processes (i.e., either correct or crashed processes depending on the failure model) can collectively explain what they observe (the crash-corrected and the mock history $H'$ are shared by all non-Byzantine processes) by assigning some ``virtual'' behavior to faulty processes: in the case of crashed processes, their virtual version might remove their last update, while for Byzantine processes, their virtual counterpart may be arbitrary.

The resulting ``virtual processes'' should be such that every correct process $p_i$ eventually sees all updates issued by virtual processes and by other non-Byzantine processes. 
Additionally, any intermediate value read by $p_i$ should be locally explainable through some serialization $S_i$ of $p_i$'s \emph{perception} of the \nameOfObjectAbbr's distributed execution (\cref{sec:histories}).
As a result, a \nameOfObjectAbbr respects both \emph{Strong Eventual Consistency} (SEC) and \emph{Pipeline Consistency} (a generalization of \emph{Pipelines Random Access Memory} (PRAM) consistency to sequential objects~\cite{PRAM88,10.5555/3134223}). The following discusses these two claims in more detail.
\newcommand{\kev}{k_{\mathsf{ev}}}%
\begin{itemize}
\item $S_i$ provides each non-Byzantine process $p_i$ with a \nameOfObjectAbbr-legal serialization that contains the updates of all processes but removes return values seen by other processes $p_j\neq p_i$. This corresponds exactly to \emph{Pipeline Consistency} as defined in~\cite{10.5555/3134223}.
\item By default, \emph{Pipeline Consistency} is incompatible with \emph{Strong Eventual Consistency}. Because updates invoked by distinct processes commute, the case of \nameOfObjectAbbrs is however different. If all processes eventually stop issuing updates, but correct processes continue to perform query operations indefinitely, then two correct processes $p_i$ and $p_j$ will eventually only return outputs that correspond to the same Mazurkiewicz trace, i.e. from the same state.
More concretely, because in such a scenario, the overall number of updates is finite, there exists some index $\kev\in \mathbb{N}$ such that the local serializations $S_i=(e_k^i)_k$ and $S_j=(e_k^j)_k$ required by the \nameOfObjectAbbr specification only contain queries after their $\kev^{\text{th}}$ invocation. Because both serializations contain the same finite set of updates, respect the process order of these updates, and respect the \emph{Process Authorization} property (\cref{sec:name-legal-sequ}), and because updates from different processes commute (\cref{eq:indep:rek:definition}), the existence of $\kev$ leads to the following trace equalities:
\begin{equation}
\forall k>\kev:
\big[\proj{(S_i)_{<k}}{\Sigma}\big]_I=
\big[\proj{S_i}{\Sigma}\big]_I=
\big[\proj{S_j}{\Sigma}\big]_I=
\big[\proj{(S_j)_{<k}}{\Sigma}\big]_I.
\end{equation}
By the \emph{Output and Query Validity} property, this implies that any query performed by $p_i$ or $p_j$ after $\kev$ will be based on the same trace/state, thus ensuring eventual consistency. In addition, the existence of $S_i$ and $S_j$ and \emph{Output and Query Validity} mean that $p_i$ and $p_j$ observe the effect of updates as soon as these become visible in their local serialization, thus fulfilling SEC as well~\cite{10.5555/3134223}.%
\footnote{\emph{Strong Eventual Consistency} (SEC)~\cite{10.5555/3134223,shapiro:inria-00555588,DBLP:journals/csur/ViottiV16} was initially defined by reference to implementation messages, which are ignored here on purpose to separate the specification of \nameOfObjectAbbr from any implementation choices. The above discussion uses therefore the version of SEC proposed by Perrin~\cite{10.5555/3134223} that eschews messages altogether.}
\end{itemize}

\section{A Generic \nameOfObjectHyph Algorithm}
\label{sec:gener-name-algor}

Given a \nameOfObjectAbbr specification $\big(C,(O_j)_j,L,Q\big)$, this section presents a simple and concise algorithm that implements the corresponding \nameOfObjectAbbr in an asynchronous message passing system. This algorithm is generic in two important ways:

\begin{itemize}

\item It works with any \nameOfObjectAbbr specification $\big(C,(O_j)_j,L,Q\big)$ (thus covering a wide range of relevant objects, including multisets, distributed Petri nets, cryptocurrencies and work-stealing objects, cf. \cref{sec:examples}.)

\item It can be instantiated with different reliable broadcasts, from which it directly inherits its fault-tolerance assumptions and properties. If the broadcast is reliable in $\CAMP[\emptyset]$, then our algorithm implements a \nameOfObjectAbbr in $\CAMP[\emptyset]$, according to the definition of \cref{sec:PCO-CAMP}. If used with a Byzantine reliable broadcast in $\BAMP[t < n/3]$, then our algorithm implements a \nameOfObjectAbbr in $\BAMP[t < n/3]$, again according to the definition in~\cref{sec:PCO-BAMP}.

\end{itemize}

From a computability point of view, we thus define and implement a family of objects that can be realized in $\CAMP[\emptyset]$, i.e., are implementable in asynchronous message-passing systems suffering from an arbitrary number of crashes (up to $n$). These objects are thus computationally strictly weaker than atomic RW memory registers (which require that at most a minority of processes may crash $t<\frac{n}{2}$), and yet encompass several useful data structures (\cref{sec:examples}).

\subsection{Reliable broadcast}
\label{sec:reliable-broadcast}

Following~\cite{auvolat:eatcs}, the proposed algorithm (whose pseudo-code is shown in \cref{algo:generic}) uses a multi-shot \emph{reliable broadcast} abstraction with explicit sequence numbers, whose concrete implementation depends on the failure model. A reliable broadcast provides the operation noted $\rbroadcast()$ (used to broadcast messages) and the callback $\rdelivered()$ (that is triggered when delivering a message). When a process $p_i$ invokes $\rbroadcast(\sn, m)$, we say ``process $i$ r-broadcasts the message $m$ with sequence number $\sn$'', and when a process receives the broadcast $\langle \sn, m \rangle$ from $p_i$, it raises the callback $\rdelivered()$.
When this happens, we say the receiving process ``r-delivers $m$ with sequence number $\sn$ from process $i$''. The exact properties of the broadcast we use depend on the failure model we assume ($\CAMP[\emptyset]$ or $\BAMP[t < n/3]$), and we return to these properties in detail in Sections~\ref{sec:camp-model} and \ref{sec:bamp-model}. We list below some of the properties shared by these two broadcast abstractions:
	
\begin{itemize}
\item {\bf Validity.}  This property states that there is no message creation.
To this end, it relates the outputs (r-deliveries) to the inputs (r-broadcasts). Excluding malicious behaviors, a message that is r-delivered has been r-broadcast.
\item {\bf Integrity.} This property states that there is no message duplication.  
\item {\bf Termination-1.} This property states that correct processes r-deliver what they broadcast.
\item {\bf Termination-2.} This property relates the sets of messages r-delivered by different processes.
\end{itemize}

The Termination properties ensure that all the correct processes r-deliver the same set of messages, and that this set includes at least all the messages that they r-broadcast.

\subsection{Local data structure}
All processes know the specification $\big(C,(O_j)_j,L,Q\big)$ of the \nameOfObject.  Each process further maintain the following local variables (\cref{line:init} of \cref{algo:generic}):
\begin{itemize}
\item $seq_i\in \Sigma^{\ast}$: the sequence of update operations processed so far by process $p_i$ (initialized to~$\epsilon$).
\item $\sn_i\in\mathbb{N}$: the sequence number of the last message broadcast by $p_i$ (initialized to $0$);
\item $del_i[1 .. n]\in\mathbb{N}^n$: a local array of the sequence numbers of the last broadcast of each process that $p_i$ has so far processed (initialized to $[0, ..., 0]$);
\end{itemize}

\newcommand{\recordLineNumber}[1]{%
  \newcounter{#1}
  \setcounter{#1}{\value{AlgoLine}}%
  \addtocounter{#1}{-1}%
}%

\subsection{Algorithm}
\label{sec:algorithm}

\newcommand{\applySc}{{\textsc{apply}}}
\newcommand{\applyFunc}{{\sf apply}}

\begin{algorithm}[tb]
\KwInit{$seq_i \leftarrow \epsilon$ ;
$\sn_i \leftarrow 0$ ; $\del_i[1..n] \leftarrow [0, \dots , 0]$.\Comment*[f]{$\epsilon$ is the empty string.}\label{line:init}}

\AlgoSkip

\Function(\Comment*[f]{Invoking the query operation
$q$}){$\applyFunc(q\in Q)$\label{line:apply:query:start}}{
\KwRet{$q(seq_i)$}\label{line:apply:query:end}\Comment*[f]{$seq_i$
captures $p_i$'s current knowledge of the \nameOfObjectAbbr's
state} }

\AlgoSkip

\Function(\Comment*[f]{Invoking the update operation $\update$}){$\applyFunc(\update\in \Sigma)$\label{line:apply:update:start}}{
\eIf(\Comment*[f]{\update{} must be authorized to $p_i$ and legal}){$\update \in C \cup O_i \wedge [seq_i \cat \update]_I \in L$ \label{line:apply:update:test}\recordLineNumber{value:apply:update:test}}{
$\sn_i \leftarrow \sn_i +1$\Comment*{$\sn_i$\textsuperscript{th} operation invoked by $p_i$}
$\rbroadcast \langle\sn_i,\applySc(\update)\rangle$\label{line:apply:rbcast}\;
\wwait{$\del_i[i]=\sn_i$}\label{line:apply:update:wait}\Comment*{Waiting for $p_i$ to process its own broadcast at \crefrange{line:rdlivery:start}{line:rdlivery:end}}
\Return{$\outputFunc(\update{},seq_i)$}
}{
\Raise{\abort}\Comment*{Should not happen if application respects condition of~\cref{line:apply:update:test}}
}
\label{line:apply:update:end}}

\AlgoSkip

\WhenDo{$\langle\sn,\applySc(\update)\rangle$ {\bf is} $\rdelivered$
  {\bf from} $p_j$\label{line:rdlivery:start}}{
  \wwait{$(\update \in C \cup O_j \wedge\sn=del_i[j]+1 \wedge [seq_i \cat \update]_I \in L )$\label{line:rdelivery:test}\recordLineNumber{value:deliver:update:seqi}}\;
  $seq_i \leftarrow seq_i \cat \update$\label{line:deliver:update:seqi}\;
  $del_i[j]\leftarrow \sn$\label{line:deliver:update:count:number}\label{line:rdlivery:end}\;
}
	\caption{A generic algorithm that implements any \nameOfObject specified by the trace language $L\subseteq\mathbb{M}(\Sigma,I)$ over the set of updates $\Sigma= C \cup \bigcup_{p_i\in \Pi}O_i$, and the set of queries $Q$. The fault-tolerance properties of the algorithm are inherited from the underlying broadcast algorithm used to implement $\rbroadcast$ (\cref{line:apply:rbcast}) and the call-back $\rdelivered$ (\cref{line:rdlivery:start}). (code for $p_i$)}
	\label{algo:generic}
	\end{algorithm}

\cref{algo:generic} provides two \applyFunc{} functions, one for query operations
(lines~\ref{line:apply:query:start}-\ref{line:apply:query:end}), and one for updates (lines~\ref{line:apply:update:start}-\ref{line:apply:update:end}).
A query operation  $q\in Q$ is purely local and just returns the result of the corresponding output $q(seq_i)$ based on $p_i$'s current sequence of observed update operations $seq_i$ (\cref{line:apply:query:end}).

When $p_i$ invokes \applyFunc{} for an update operation $\update$ (for short we will say that $p_i$ invokes $\update$), this operation must be authorized for $p_i$ and legal (\cref{line:apply:update:test}). Assuming it is, $p_i$ computes its next sequence number $\sn_i$ and r-broadcasts the pair $\langle \sn_i,\applySc(\update)\rangle$. The predicate at \cref{line:apply:update:wait} is satisfied when $p_i$ r-delivers its own broadcast at \cref{line:deliver:update:count:number}, at which point the function returns.

When $p_i$ r-delivers a pair $\langle \sn,\applySc(\update) \rangle$ from $p_j$, it waits until it has processed each previous message sent by $p_j$ (FIFO order) and until $\update$ is legal when concatenated with $seq_i$, i.e. until $[seq_i \cat \update]_I \in L$ (\cref{line:rdelivery:test}). Once those two conditions are satisfied, $p_i$ updates its sequence of operations and increases $del_i[j]$.

\subsection{Practical implementation}

\cref{algo:generic} uses the trace language $L$ and the independence relation $\sim_I$ in line with the presentation of \nameOfObject in \cref{sec:nameofobjects}. Traces nicely capture the partial commutativity of some updates through the algebraic structure of the monoid $\mathbb{M}(\Sigma,I)$, and offer an elegant framework to reason about distribution and concurrency.

Implementing a \nameOfObjectAbbr using a trace language would, however, be impractical. Instead, given a \nameOfObjectAbbr specification $\big(C,(O_j)_{j\in[1..n]},L,Q\big)$, the structure of $L$ can be converted to a state automaton using an appropriate equivalence relation on the traces of $L$. This construction is detailed in \cref{app:conv-name-spec}.

\section{Proofs}
\label{sec:proofs}

This section first considers crashes ($\CAMP[\emptyset]$) and then Byzantine processes ($\BAMP[t < n/3]$). In each system model it states the formal guarantees provided by the multi-shot reliable broadcast abstraction, noted \emph{CR-Broadcast} (CRB for short) in $\CAMP[\emptyset]$, and \emph{BR-Broadcast} (BRB for short) in $\BAMP[t < n/3]$. Building upon these definitions, the section proves the following two results regarding \cref{algo:generic}.

\begin{restatable}[\nameOfObjectAbbr with crashes]{theorem}{theoCrashProof}\label{crash-proof}
\cref{algo:generic} instantiated with CR-Broadcast implements a \nameOfObject specified by the tuple $\big(C,(O_j)_j,L,Q\big)$ in the system model $\CAMP[\emptyset]$.
\end{restatable}

\begin{restatable}[\nameOfObjectAbbr with Byzantine failures]{theorem}{theoByzantine}\label{byzantine-proof}
\cref{algo:generic} instantiated with BR-Broadcast implements a \nameOfObject specified by the tuple $\big(C,(O_j)_j,L,Q\big)$ in the system model $\BAMP[t < n/3]$.
\end{restatable}

\subsection{Notation and definitions}
\begin{itemize}

\item $\update^{\sn}_j$ denotes the update operation issued by $p_j$ with sequence number $\sn$.

\item We say a process $p_i$ \emph{processes} the update operation $\update^{\sn}_j$ if, after it r-delivered the associated message $\langle \sn, \applySc(\update^{sn}_j)\rangle$ at \cref{line:rdlivery:start}, $p_i$ exits the wait statement at \cref{line:rdelivery:test} and executes the associated statements at \cref{line:deliver:update:seqi,line:deliver:update:count:number}.

\item We say that an update $\update^{\sn}_j$ is \emph{successful} if the message $\langle \sn, \applySc(\update^{\sn}_j)\rangle$ r-broadcast by $p_j$ (correct or not) is processed by at least one correct process.

\end{itemize}

\subsection{In the $\CAMP[\emptyset]$ model}
\label{sec:camp-model}

\subsubsection{Multi-shot reliable broadcast abstraction in the crash failure model}

Crash-tolerant Reliable Broadcast (CR-Broadcast, CRB for short) refines in a crash failure model the properties presented in \cref{sec:reliable-broadcast}. CR-Broadcast provides the operation ${\sf{cr\_broadcast()}}$ and the callback ${\sf{cr\_delivered()}}$, which refine $\rbroadcast()$ and $\rdelivered()$ introduced in \cref{sec:reliable-broadcast}. In the following, we use the terminology ``to cr-broadcast a message'', and ``to cr-deliver a message''. CR-Broadcast is defined by the following properties~\cite{raynal-book18}.

\begin{itemize}
\item CRB-Validity: If a process $p_i$ cr-delivers a message with sequence number $\sn$ from a process $p_j$, then $p_j$ cr-broadcast it with sequence number $\sn$.
\item CRB-Integrity: For each sequence number $\sn$ and sender $p_j$ a process $p_i$ cr-delivers at most one message with sequence $\sn$ from $p_j$.
\item CRB-Termination-1: If a correct process cr-broadcasts a message, it	cr-delivers it.
\item CRB-Termination-2: If a process cr-delivers a message from a (correct or	faulty) process $p_j$, then all correct processes cr-deliver it.
\end{itemize}

\subsubsection{Proof of \cref{crash-proof}}

\begin{lemma}\label{lemma:crash-termination}
All invocations of update and query operations issued by a correct process terminate.
\end{lemma}

\begin{proofL}
The fact that query operations terminate follows immediately from the pseudo-code of \cref{algo:generic} (\cref{line:apply:query:start,line:apply:query:end}).

When a correct process $p_i$ invokes an update operation $\update$, it cr-broadcasts the message $\langle \sn_i,\applySc(\update) \rangle$. Because of the CRB-Termination-1 property, $p_i$ eventually cr-delivers its own message. By construction of the algorithm, $\sn_i$ counts the number of r-broadcasts performed by $p_i$, while $del_i[i]$ counts the number of its own broadcasts that $p_i$ has already processed. The \wwait{} statement at \cref{line:apply:update:wait}, implies that when $p_i$ invokes $\update$ all earlier updates invoked by $p_i$ have been processed, and no new update can be invoked (and hence processed, by CRB-Validity), until $p_i$ has processed $\update$, which can occur only once by CRB-Integrity. As a result, the condition $\sn = del_i[i] + 1$ is verified at \cref{line:rdelivery:test} when $p_i$ cr-delivers the message corresponding to $\update$.

Turning to the second part of the condition of \cref{line:rdelivery:test}, the only updates that $p_i$ can process between the broadcast of $\update$ and its crb-delivery by $p_i$ are updates by other processes. As a result, if we note $seq_i^{\mathrm{bcast}}$ the value of $seq_i$ when $\update$ is crb-broadcast at \cref{line:apply:rbcast}, and $seq_i^{\mathrm{deli}}$ its value when $\update$ is crb-delivered at \cref{line:rdelivery:test}, then $seq_i^{\mathrm{deli}}=seq_i^{\mathrm{bcast}}\cat z$ with $z\in (\Sigma\setminus O_i)^{\ast}$. Because of \cref{line:apply:update:test}, we have $[seq_i^{\mathrm{bcast}}\cat \update]_I\in L$, and because of the second part of the condition at \cref{line:rdelivery:test}, we have $seq_i^{\mathrm{deli}}=seq_i^{\mathrm{bcast}}\cat z\in L$. Using the $I$-diamond Closure of $L$, we get that $[seq_i^{\mathrm{bcast}}\cat z \cat \update]_I\in L$, and hence that $[seq_i^{\mathrm{deli}}\cat \update]_I\in L$, which concludes the proof.
\renewcommand{\toto}{lemma:crash-termination}
\end{proofL}

\begin{corollary}\label{corollary:crash-all-correct-updates-successful}
  All updates invoked by correct processes are successful.
\end{corollary}

\begin{proofC}
  This directly follows from \cref{lemma:crash-termination}.
\renewcommand{\toto}{corollary:crash-all-correct-updates-successful}
\end{proofC}

\begin{lemma}\label{lemma:crash-termination2}
If a process $p_i$ processes $\update_\ell^{\sn}$, then all correct processes process $\update_\ell^{\sn}$.
\end{lemma}

\begin{proofL}
We note $seq_i^k=m_1 \cat \cdots \cat m_{k-1} \cat m_{k}$ the prefix of $seq_i$ of size $k$. We show by induction on $k$, that for every correct process $p_j$, and for every integer $k\leq |seq_i|$~\footnote{By convention, we assume $|seq_i|=\infty$ when $seq_i$ is infinite, and $\infty-1=\infty$ just below.}, $seq_i^k$ eventually becomes a trace prefix of $seq_j$ (i.e. $[seq_j]_I=[seq_i^k]_I\cat w$ for some trace $w$), which implies that $p_j$ eventually processes all operations of $seq_i^k$.

\begin{itemize}

\item Case $k=0$, trivial

\item $0 \leq k \leq |seq_i|-1$, we suppose the induction property true for $k$. Let us consider $p_j$ a correct process. We first prove by contradiction that (A) $p_j$ eventually processes $m_{k+1}$, and then show this implies that (B) the property is true for $k+1$.

\emph{Part A:} Let us assume that $p_j$ never processes $m_{k+1}$. By induction hypothesis, at some point we have $[seq_j]_I=[seq_i^k]_I\cat w$ for some $w$, and $p_j$ has processed $m_1,...m_k$. Because of the CRB-Termination-2 property, $p_j$ eventually cr-delivers $m_{k+1}$. As $p_j$ never processes $m_{k+1}$, this means the condition at \cref{line:rdelivery:test} never becomes true. Let us assume $m_{k+1} = \update_\ell^{\sn}$, i.e. $m_{k+1}$ is the $\sn$\textsuperscript{th} update invoked by some process $p_\ell$. Consider a time point after $\update_\ell^{\sn}$ has been cr-delivered to $p_i$ and $p_j$, and after $[seq_j]_I=[seq_i^k]_I\cat w$ for some $w$. At that point, we have

\begin{itemize}

\item Since the messages cr-broadcast by one process are processed in FIFO order (as a result of CRB-Integrity and the use of $del_i[j]$ at \cref{line:rdelivery:test,line:deliver:update:count:number}), the fact that $\update_\ell^{\sn}$ is never processed by $p_j$ implies $del_j[\ell] \leq \sn-1$. 
If $\sn=1$, as $del_j[\ell]\geq 0$, this yields $del_j[\ell]=0=\sn-1$.
If $\sn>1$, as $\update_\ell^{\sn}$ has been processed by $p_i$, $\update_\ell^{\sn-1}$ has been processed as well by $p_i$ and therefore $\update_\ell^{\sn-1} \in \{m_1\cdots m_{k}\}$. By our induction hypothesis, $\update_\ell^{\sn-1}$ has been processed by $p_j$ and therefore $del_j[\ell] \geq \sn -1$. We conclude that in both cases $del_j[\ell] = \sn -1$.

\item At this same instant, we know that \begin{equation}[seq_j]_I \in L\label{eq:seqj:in:L}\end{equation} (because of the satement at \cref{line:rdelivery:test}, we only add operations that preserve $seq_j$'s legality.) If $\update_\ell^{\sn} \in C$, as $LC \subseteq L$ (by $C^{\ast}$-Closure), we have $[seq_j \cat \update_\ell^{\sn}]_I \in L$, rendering the condition of \cref{line:rdelivery:test} true, and contradicting the fact that $\update_\ell^{\sn}$ is never processed. This implies that $\update_\ell^{\sn} \in O_\ell$. Because $p_i$ processes $\update_\ell^{\sn}$, we have
  \begin{equation}
    [seq_i^{k}]_I\cat \update_\ell^{\sn}=[seq_i^{k}\cat \update_\ell^{\sn}]_I=[seq_i^{k}\cat m_{k+1}]_I=[seq_i^{k+1}]_I
    \in L,\label{eq:seqik:opl:in:L}
  \end{equation}
due to the condition at \cref{line:rdelivery:test} when $p_i$ processes $\update_\ell^{\sn}$.
By induction hypothesis, we also have \begin{equation}[seq_j]_I=[seq_i^k]_I\cat w\label{eq:seqj:seqik:w},\end{equation} for some $w$. Observing that the number of operations from $p_\ell$ in both $seq_i^k$ and $seq_j$ is exactly $\sn-1$, we infer that $w$ does not contain any operation from $p_\ell$: \begin{equation}w\in \mathbb{M}(\Sigma\setminus O_\ell).\label{eq:w:no:ol}\end{equation} Combining Equations~\ref{eq:seqj:in:L} to~\ref{eq:w:no:ol}, we can apply the $I$-diamond closure property of $L$, and conclude that $[seq_j]_I\cat \update_\ell^{\sn} = [seq_i^k]_I\cat w \cat \update_\ell^{\sn} \in L$, which means the second conditions at \cref{line:rdelivery:test} becomes true, that $\update_\ell^{\sn}=m_{k+1}$ is eventually processed by $p_j$, concluding Part A of the proof.
\end{itemize}

\emph{Part B:} We know that $p_j$ eventually processes $\update_\ell^{\sn}=m_{k+1}$, and by induction hypothesis that eventually $[seq_j]_I=[seq_i^k]_I\cat w$ for some trace $w$. Let us consider a time point when both statements have become true. As $seq_i^k$ already contains the $\sn-1$ first operations of $p_\ell$, $\update_\ell^{\sn}$ must appear in $w$, and we therefore have $[seq_j]_I=[seq_i^k]_I\cat w_1\cat \update_\ell^{\sn}\cat w_2$ for some traces $w_1$ and $w_2$, with $w_1 \in \mathbb{M}(\Sigma\setminus O_\ell)$. $\update_\ell^{\sn}$ therefore commutes with $w_1$, and we have $[seq_j]_I=[seq_i^k]_I\cat \update_\ell^{\sn} \cat w_1 \cat w_2=[seq_i^{k+1}]_I \cat w'$, with $w'=w_1\cat w_2$, which proves the property for $k+1$.%
\end{itemize}\vspace{-\baselineskip}%
\renewcommand{\toto}{lemma:crash-termination2}%
\end{proofL}

\begin{corollary}\label{corollary:crash-termination3}
A successful update operation $\update_\ell^{\sn}$ is eventually processed by all correct processes.
\end{corollary}

\begin{proofC}
Since the operation is successful, it is processed by at least one correct process, implying by \cref{lemma:crash-termination2}, that all correct processes eventually process the operation. \renewcommand{\toto}{corollary:crash-termination3}%
\end{proofC}

\theoCrashProof*

\begin{proofT}
Lemma~\ref{lemma:crash-termination} proved that all the operations invoked by correct processes terminate. To complete the proof, we need to show the \nameOfObjectAbbr-compliance in the crash failure model (\cref{def:PCO:compliance:CAMP},  \cref{sec:PCO-CAMP}). Consider an execution of \cref{algo:generic}, captured by the history $H=(E_1,\dots, E_n)$.

We choose our crash-corrected history $H' = (E'_1, \dots, E'_n)$ such that
\begin{itemize}
\item if $p_k$ only issues successful updates (this is the case if $p_k$ correct), $E'_k = E_k$;
\item if $p_k$ is faulty and the last update it invokes is not successful (this happens when $p_k$ crashes while executing
\cref{line:apply:rbcast}; all earlier updates must be successful due to the wait statement at \cref{line:apply:update:wait}), we pick $E'_k$ as $E_k$ without its last element (which is the unsuccessful update operation).
\end{itemize}
Let us consider a process $p_i$, that might be correct of faulty. Let $S_i^{\triangleleft}$ be the sequence of invocations $(\op, \val, p_i)$ that contains \emph{(i)} the query operations $(q, \val, p_i)\in Q\times\mathbb{V}\times\{p_i\}$ invoked by $p_i$, and \emph{(ii)} the update operations $(\update, \val', p_j)\in \Sigma\times\big(\mathbb{V}\cup\{\unknown\}\big)\times\Pi$ processed by $p_i$, whose output is masked when $p_j\neq p_i$ (\cref{sec:histories}). I.e., if $(\update, \val, p_j)$ is the actual update processed by $p_i$, $\val'$ is $\unknown$ if $j\neq i$ and $\val$ otherwise. The invocations of  $S_i^{\triangleleft}$ follow the order in which these queries and updates occurred at $p_i$ (i.e. the order in which the lines~\ref{line:apply:query:end} and~\ref{line:deliver:update:seqi} were executed).
\begin{itemize}
\item 
If $p_i$ is correct, we define $S_i$ as $S_i=S_i^{\triangleleft}$.
\item
  If $p_i$ is faulty, we consider an equivalent virtual correct process $p'_i$ that behaves as $p_i$ until it crashes and then remains quiet indefinitely, i.e., $p'_i$ only processes successful updates that $p_i$ did not process, but does not invoke any query or update. $p'_i$ is indistinguishable from $p_i$ from the point of view of other processes. Since it behaves as a correct process, because of Lemma~\ref{corollary:crash-termination3}, it eventually processes all successful updates that the original faulty process $p_i$ did not process. We define $S_i^{\triangleright}$ as the sequence of successful updates processed by $p'_i$ but not by $p_i$, with their output masked using the special value $\unknown$, and define $S_i$ as $S_i=S_i^{\triangleleft}\cat S_i^{\triangleright}$.
\end{itemize}
We first prove $S_i$ is a serialization of $\widehat{H_i'}$ ($p_i$'s perceived history of $H'$, defined in \cref{sec:histories}).
\begin{itemize}

\item All invocations of $S_i$ appear in $\widehat{H_i'}=(\widehat{E'}_1, \dots, \widehat{E'}_n)$: if $(q, \val, p_i)\in S_i$ is a query, $q$ was invoked by $p_i$ and so is part of $E'_i=\widehat{E'}_i$. If $(\update, -, p_\ell)\in S_i$ is an update, we have two cases. Case 1, if $(\update, -, p_\ell)\in S_i^{\triangleleft}$, then $\update$ was processed by $p_i$. By CRB-Validity, $\update$ was issued by $p_\ell$, and by \cref{lemma:crash-termination2} it corresponds to a successful update of $p_\ell$. As a result, $(\update, -, p_\ell)\in \widehat{E'}_\ell$. Case 2, if $(\update, -, p_\ell)\in S_i^{\triangleright}$, by construction $(\update, -, p_\ell)$ was successful and similarly $(\update, -, p_\ell)\in \widehat{E'}_\ell$.

\item All invocations in $\widehat{H_i'}=(\widehat{E'}_1, \dots, \widehat{E'}_n)$ appear in $S_i$: for $p_\ell\neq p_i$, $\widehat{E'}_\ell$ only contains successful update operations (since all updates issued by a correct process are successful by \cref{corollary:crash-all-correct-updates-successful}, and the only update issued by a faulty process that might be unsuccessful---the last one---is removed from $\widehat{E'}_\ell$). By construction, all these updates are present in $S_i$: due to \cref{corollary:crash-termination3} if $p_i$ is correct, and by construction of $S_i^{\triangleleft}$ and $S_i^{\triangleright}$ if $p_i$ is faulty. Turning to $p_i$'s execution, $\widehat{E'}_i$ contains all the queries invoked by $p_i$ and all successful updates issued by $p_i$. By definition, all queries invoked by $p_i$ are in $S_i$. The reasoning on the updates of $\widehat{E'}_i$ is the same as for $\widehat{E'}_{\ell\neq i}$.

\item The total order of $S_i$ preserves the process orders $(\rightarrow_i)_i$ of $\widehat{H_i'}$. Consider a process $p_j$ that invokes two update operations, $\update_1$ and $\update_2$, with sequence numbers $\sn_1$ and $\sn_2$ respectively. If $\update_1$ appears before $\update_2$ in $\widehat{E'}_j$, $p_j$ must have issued $\update_1$ before $\update_2$, therefore, $\sn_1 < \sn_2$. The condition on $\sn$ at \cref{line:rdelivery:test} ensures that $p_i$ does not process $\update_2$ before $\update_1$. If $p_i$ is faulty and crashes before processing any of the updates, or processes $\update_1$ and crashes before processing $\update_2$, the construction of $S_i^{\triangleright}$, and the fact that $S_i^{\triangleright}$ appears after $S_i^{\triangleleft}$ in $S_i$ ensures the order between $\update_1$ and $\update_2$ is respected in $S_i$. In $\widehat{E_i}$, if an update operation $\update$ is invoked before a query $q$, \cref{line:apply:update:wait} insures
that $\update$ is processed before $q$ in $S_i$. The reverse (when $q$ is invoked before $\update$) is also trivially true. As a result, the order between queries and updates in $\widehat{E_i}$ is also preserved in $S_i$.
\end{itemize}
Now that we have proved that $S_i$ is a serialization of $\widehat{H_i}$, let us show that $S_i$ is \nameOfObjectAbbr-legal (\cref{sec:name-legal-sequ}).
\begin{itemize}
\item \emph{Process Authorization} is guaranteed by the condition $\update \in C \cup O_j$ of \cref{line:rdelivery:test}.

\item \emph{Update Legality.} Let us consider a prefix $u_i$ of $\proj{S_i}{\Sigma}$. If $p_i$ is correct, by construction of $S_i$, there exists a point in the execution of $p_i$ when $u_i = seq_i$. Because of \cref{line:rdelivery:test}, $[seq_i]_I\in L$, and therefore $[u_i]_I \in L$. If $p_i$ is faulty, $u_i$ is such that $u_i = seq'_i$ at some point of the execution of $p'_i$, where $seq'_i$ is the sequence of updates constructed by $p'_i$, and $p'_i$ is the virtual correct process corresponding to $p_i$ until its crash, used to construct $S_i$. By the same argument as for correct processes, $[u_i]_I \in L$.

\item \emph{Output and Query Validity.} The only query operations present in $S_i$ are those invoked by $p_i$, by construction. When a query operation $q$ is invoked by $p_i$, \cref{algo:generic} returns $q(seq_i^q)$ where $seq_i^q$ denotes the sequence of update operations $p_i$ has processed until the query $q$. If we note $e_k = (q, \val, p)$, then $\val = q(seq_i^q)$. By construction of $S_i$, $seq_i^q=\proj{S_i^{< k}}{\Sigma}$, where $S_i^{< k}$ represents all invocations which precede $e_k$ in $S_i$, and therefore $\val = q\big(\proj{S_i^{< k}}{\Sigma}\big)$.

By construction, the only updates $e_k=(\update,val\neq \unknown,-)$ of $S_i$ associated with a visible value are those performed by $p_i$. A reasoning similar to that for queries leads to $\val =\outputFunc\big(\update,[\proj{S_i{<k}}{\Sigma}]_I\big)$, concluding the proof.
\end{itemize}
\vspace{-\baselineskip}~%
\renewcommand{\toto}{crash-proof}%
\end{proofT}

	\subsection{In the $\BAMP[t < n/3]$ model}
\label{sec:bamp-model}

\subsubsection{Reliable broadcast abstraction in the Byzantine failure model}

Byzantine Tolerant Reliable Broadcast (BR-Broadcast, BRB for short) instantiates the reliable broadcast abstraction of \cref{sec:reliable-broadcast} in $\BAMP[t < n/3]$. It provides the operation ${\sf{br\_broadcast()}}$ and the callback ${\sf{br\_delivered()}}$, which refine $\rbroadcast()$ and $\rdelivered()$ presented in \cref{sec:reliable-broadcast}. As for CR-Broadcast, we use the terminology ``to br-broadcast a message'' and ``to br-deliver a
message'' in the following. BR-Broadcast is defined by the following properties~\cite{raynal-book18,LSP82,bracha1987asynchronous}.

\begin{itemize}
\item BRB-Validity.  If a \emph{correct} process $p_i$ br-delivers a
message from a \emph{correct} $p_j$ with sequence number $\sn$, then
$p_j$ br-broadcast it with sequence number $\sn$.
		\item  BRB-Integrity.
		For each sequence number $\sn$ and sender $p_j$ a \emph{correct} process $p_i$ br-delivers at most one message with	sequence number $\sn$ from sender $p_j$.
		\item BRB-Termination-1. 
    If a correct	process br-broadcasts a message, it br-delivers it.
		\item BRB-Termination-2.
    If a \emph{correct} process  br-delivers a message from a (correct or faulty) process $p_j$, then all correct processes br-deliver it.
\end{itemize}

\subsubsection{Proof of \cref{byzantine-proof}}

\paragraph{Notation} If $p_j$ is Byzantine, $\update^{\sn}_j$ does not necessarily correspond to an actual update invoked by $p_j$. However, the BRB-Termination-2 property guarantees that correct processes all br-deliver at \cref{line:rdlivery:start} the \emph{same} set of update messages $\langle\sn,\applySc(\update)\rangle$ from $p_j$, and therefore agree on how $p_j$'s behavior should be interpreted.

\cref{lemma:crash-termination,corollary:crash-all-correct-updates-successful,lemma:crash-termination2,corollary:crash-termination3} are still valid in this model, by considering only \emph{correct} processes. This observation yields the three following lemmas when \cref{algo:generic} is executed in $\BAMP[t<n/3]$: 

	\begin{lemma}\label{lemma:byzantine-termination}
	  All invocations of update and query operations issued by a correct process terminate.
	\end{lemma}

  \begin{corollary}\label{corollary:byzantine-all-correct-updates-successful}
    All updates invoked by correct processes are successful.
  \end{corollary}

	\begin{lemma}\label{lemma:byzantine-termination2}
		If a correct process $p_i$ processes $\update_\ell^{\sn}$, then all correct processes process $\update_\ell^{\sn}$.
	\end{lemma}

	\begin{corollary}\label{corollary:byzantine-termination3}
    A successful update operation $\update_\ell^{\sn}$ is eventually processed by all correct processes.
	\end{corollary}

	The proofs of those three lemmas are analog to proofs of \cref{lemma:crash-termination,corollary:crash-all-correct-updates-successful,lemma:crash-termination2,corollary:crash-termination3}, by replacing the use of CRB-* properties by their corresponding BRB-* ones.

\theoByzantine*

\begin{proofT}
The constraint $t < n /3$ is due to the BR-broadcast. 
Let us first note that termination is provided by Lemma~\ref{lemma:byzantine-termination}.

We consider an execution of the algorithm, captured as a history $H
= (E_1, \dots, E_n)$.  We define the mock local history $H' = (E'_1,
\dots, E'_n)$ in the following manner:
\begin{itemize}
\item If $p_i$ is correct, $E'_i=E_i$.
\item If $p_i$ is Byzantine, $E'_i$ is the set of update operations $(\update_i^{\sn})_{\sn}$ \brdeliveredtext from $p_i$ and processed by at least one correct process, ordered according to their sequence numbers. Because of Lemma~\ref{lemma:byzantine-termination2}, all correct processes process the updates of $E'_i$, and due to the conditions $\sn=del_i[j]+1$ at \cref{line:rdelivery:test}, they have all processed these updates in the order in which they appear in $E'_i$.
\end{itemize}

Let $p_i$ be a correct process. Let $S_i$ be the sequence of invocations $(\op, \val, p_i)$ that contains \emph{(i)} the query operations $(q, \val, p_i)\in Q\times\mathbb{V}\times\{p_i\}$ invoked by $p_i$, and \emph{(ii)} the update operations $(\update, \val', p_j)\in \Sigma\times\big(\mathbb{V}\cup\{\unknown\}\big)\times\Pi$ processed by $p_i$, whose output is masked when $p_j\neq p_i$ (\cref{sec:histories}). I.e., if $(\update, \val, p_j)$ is the actual update processed by $p_i$, $\val'$ is $\unknown$ if $j\neq i$ and $\val$ otherwise. The invocations of  $S_i$ follow the order in which these queries and updates occurred at $p_i$ (i.e., the order in which the lines~\ref{line:apply:query:end} and~\ref{line:deliver:update:seqi} were executed).

We first prove that $S_i$ is a serialization of $\widehat{H_i'}$.
\begin{itemize}
  
\item All invocations of $S_i$ appear in $\widehat{H_i'}=(\widehat{E'}_1, \dots, \widehat{E'}_n)$: if $(q, \val, p_i)\in S_i$ is a query, $q$ was invoked by $p_i$ and so is part of $E_i=E'_i=\widehat{E'}_i$ (since $p_i$ is correct). If $(\update, -, p_\ell)\in S_i$ is an update, we have two cases. Case 1, if $p_\ell$ is correct, then by BRB-Validity, $\update$ was issued by $p_\ell$, and $(\update, -, p_\ell)\in \widehat{E'}_\ell$. Case 2, if $p_\ell$ is Byzantine, then since $p_i$ (which is correct) processed $(\update, -, p_\ell)$, then we have $(\update, -, p_\ell)\in E'_\ell = \widehat{E'}_\ell$.

\item All invocations in $\widehat{H_i'}$ appear in $S_i$: first consider a process $p_\ell\neq p_i$. If $p_\ell$ is correct, $\widehat{E'}_\ell$ only contains update operations issued by $p_\ell$. By Lemma~\ref{corollary:byzantine-termination3}, $p_i$ will eventually process these updates, which will therefore appear in $S_i$. If $p_\ell$ is Byzantine, by construction, $\widehat{E'}_\ell$ only contains update operations processed by all correct processes, and hence by $p_i$ and these operations will also appear in $S_i$. $\widehat{E'}_i$ contains all the queries invoked by $p_i$ and all updates issued by $p_i$. By definition, all queries invoked by $p_i$ are in $S_i$. The reasoning on the updates of $\widehat{E'}_i$ is the same as for $\widehat{E'}_{\ell\neq i}$.

\item The total order of $S_i$ preserves the process orders $(\rightarrow_i)_i$ of $\widehat{H_i'}$: For each process $p_j$, if an update $\update_1$ appears before $\update_2$ in $\widehat{E'}_j$, it means that $\sn_1 < \sn_2$ ($\sn_k$ being the sequence number associated with $\update_k$ during the execution), either because $p_j$ is correct, or by construction of $E'_j$ if $p_j$ is Byzantine. The condition on $\sn$ at \cref{line:rdelivery:test} ensures that $p_i$ does not process $\update_2$ before $\update_1$. In $\widehat{E_i}$, if an update operation $\update$ is invoked before a query $q$, \cref{line:apply:update:wait} insures that updates issued by $p_i$ are processed at \cref{line:deliver:update:seqi} in the order in which they are invoked at \cref{line:apply:update:start}, and that $\update$ is processed before $q$ in $S_i$. The reverse (when $q$ is invoked before $\update$) is also trivially true. As a result, the order between queries and updates in $\widehat{E_i}$ is also preserved in $S_i$.
  
\end{itemize}

The reasoning to prove that $S_i$ is \nameOfObjectAbbr-legal is similar to that of Theorem~\ref{crash-proof}. The arguments for \emph{Process Authorization} and \emph{Query Validity} are the same, and that of \emph{Update Legality} is somewhat simpler, as we only need to consider the correct case:
\begin{itemize}
\item \emph{Update Legality.} Let us consider a prefix $u_i$ of $\proj{S_i}{\Sigma}$. Since $p_i$ is correct, by construction of $S_i$, a point exists in the execution of $p_i$ when $u_i = seq_i$. Because of \cref{line:rdelivery:test}, $[seq_i]_I\in L$, and therefore $[u_i]_I \in L$.%
\end{itemize}\vspace{-\baselineskip}~%
\renewcommand{\toto}{byzantine-proof}
\end{proofT}


\section{Examples}
\label{sec:examples}

This section presents four practical examples of \nameOfObjectAbbr objects: (i)~\emph{Multi-set with Deletion Rights}, (ii)~\emph{Shared Petri Net with Transition Rights}, (iii)~\emph{Money Transfer}, and (iv)~\emph{Idempotent Work Stealing Deque}. These examples cover fundamental data structures that can be used to build more complex systems (multi-sets and Petri nets) and concrete distributed services found in actual applications (money transfer and work stealing).

For each example, we define $C$, the list of common updates; $(O_i)_i$, the list of process-specific updates; the trace language $L$ capturing the legal update sequences as well as the object's state; and finally, the set of query functions $Q=\{q_k\}_k$ over $L$ that the object implements.

In each case, we prove that the proposed language $L$ verifies the three properties introduced in Section~\ref{sec:sequ-spec} (\emph{Initial emptiness}, \emph{$C^{\ast}$-Closure}, and \emph{$I$-diamond Closure}) that are required by \cref{algo:generic} to implement a Distributed Process-Commutative Object (either in the $\CAMP[\emptyset]$ or $\BAMP[t < n/3]$ model, depending on the reliable broadcast that is used), as defined in Section~\ref{sec:name-compl}.

\subsection{Multi-set with Deletion Rights}
\label{sec:multi-set-with}
\newcommand{\add}{\mathsf{add}}
\newcommand{\delete}{\mathsf{delete}}
\newcommand{\getSet}{\mathsf{get\_set}}
\newcommand{\multipquery}{\mathit{multip}}
\newcommand{\setOfElem}{E}
Distributed sets are a canonical example of CRDTs that can be used to construct more complex applications. Because addition operations to a set commute, grow-only sets (G-Set~\cite{shapiro:inria-00555588}) are straightforward to implement as a CRDT. Deletions pose a more difficult challenge as adding and removing the same element do not commute. The classical solution consists in arbitrating between \emph{add} and \emph{remove} operations deterministically, leading to \emph{remove-wins} and \emph{add-wins} solutions (2P-Set and OR-Set respectively in~\cite{shapiro:inria-00555588}), in which one of the two operations is given precedence over the other. One drawback is, however, that these solutions no longer respect the local semantic of set operations: for instance, in a \emph{remove-wins} design, an element recorded as removed will never appear in the set again. Adding this element simply has no effect.

The following proposes an alternative design that implements a distributed multi-set as a \nameOfObjectAbbr. The multiplicity of each element makes it possible to track how many times an element has been added. As in CRDT-based sets, removals are more challenging to implement as an element that is absent (multiplicity of 0) cannot be removed. The proposed design overcomes this difficulty by allowing only one specific process to remove a given element (this process as the \emph{deletion right} for this element) and forbidding the removal of absent elements. This strategy thus preserves the semantic of addition and removal and ensures, in particular, that multiplicity values remain non-negative.
\subsubsection{Definition}

A shared multi-set object with deletion rights provides two classes of
update operations: $\add_e$ adds the element $e\in \setOfElem$ to the
multiset, while $\delete_e$ removes one instance of $e$ from the
multiset. Elements might appear multiple times in the multiset, and
can be added by any process. A given element $e$ can, however, only be
deleted by a specific process that has the \emph{deletion right} for
this element. We note $\setOfElem_i$ the set of elements that $p_i$
can delete. The sets $(\setOfElem_i)_{p_i\in\Pi}$ partition
$\setOfElem$, i.e. $\setOfElem=\bigcup_{p_i\in\Pi}\setOfElem_i$, and
$\setOfElem_i\cap\setOfElem_j=\emptyset$ for $i\neq j$. An element
must also be present as least once in the multiset to be deleted.

A shared multi-set object supports a single query operation,
$\getSet$, which returns a finite map $\multipquery\in\mathbb{Z}^{\setOfElem}$ that provides the multiplicity $\multipquery(e)$ of each element $e$ in the multiset. As we are dealing with finite mutlisets, $\multipquery$ is non-zero for only a finite number of elements\footnote{The definition of the trace language $L$ below further ensures that $\multipquery(e)$ remains non-negative. However, because the queries of \nameOfObjectAbbr are formally defined over any trace of $\mathbb{M}(\Sigma)$, the codomain of $\multipquery$ has to be set to $\mathbb{Z}$.}. If $\getSet$ is invoked after the $\add$ and $\delete$ operations contained in a trace $v$, then
\begin{equation}\label{eq:def:getSet}
\getSet(v)(e)=|\proj{v}{\add_e}|-|\proj{v}{\delete_e}|,
\end{equation}
where $\proj{v}{x}$ is the projection of the trace $v$ that only retains the operation $x$.

\subsubsection{Alphabet and language definitions}
Formally, we define the sets of $C$, $(O_i)_i$, and $Q$ of a shared multi-set with deletion rights as follows:
\begin{itemize}
	\item $C   \isdefinedas \{\add_e\}_{e\in \setOfElem}$,
  \item $O_i \isdefinedas \{\delete_e\}_{e \in \setOfElem_i}$,
	\item $Q   \isdefinedas \{\getSet\}$.
\end{itemize}
$\add_e$ and $\delete_e$ are silent update operations, i.e. their respective $\outputFunc$ functions always return some pre-determined constant value (e.g. $\bot$).

The trace language of a shared multi-set object with deletion rights is defined as
\begin{equation}\label{eq:L:shareMultiSet}
L \isdefinedas \big\{v\in \mathbb{M}(\Sigma) \mid \forall e \in \setOfElem: \getSet(v)(e)\geq 0\big\}.
\end{equation}

\begin{theorem}\label{th:multiset:PCO}
	$\big(C,(O_j)_j,L,Q\big)$ as defined above for a shared multi-set object with deletion rights specifies a \nameOfObject.
\end{theorem}

\begin{proofT}\renewcommand{\toto}{th:multiset:PCO}
  We need to prove that $L$ verifies the properties of \emph{Initial emptiness}, \emph{$C^{\ast}$-Closure}, and \emph{$I$-diamond Closure} defined in Section~\ref{sec:sequ-spec}.
  \begin{itemize}

  \item \emph{Initial emptiness:} $\getSet(\epsilon)(e)=|\proj{\epsilon}{\add_e}|-|\proj{\epsilon}{\delete_e}|=0-0$, and therefore $\epsilon\in L$.

  \item \emph{$C^{\ast}$-Closure:} Consider a trace $v \in L$, and an add operation $\add_e\in C$ on an element $e$. Since $v\in L$, for any $e'\in \setOfElem$, we have $\getSet(v)(e')\geq 0$. This observation, and the fact that $\getSet(v\cat \add_e)(e) = \getSet(v)(e) +1$ by construction, implies that $\getSet(v\cat \add_e)(e)\geq 0$. Considering now an element $e'\neq e$, we have $\getSet(v\cat \add_e)(e')=\getSet(v\cat)(e')$, also yielding $\getSet(v\cat \add_e)(e')\geq 0$ by the same reasoning.

  \item \emph{$I$-diamond Closure:} Consider a trace $v \in L$, an operation $\delete_{e_i}$ with $e_i\in \setOfElem_i$ , and a trace of updates $z \in \mathbb{M}(\Sigma \setminus O_i)$, so that $\{v\cat \delete_{e_i}, v\cat z\}\subseteq L$. We need to consider the case of $e_i$ and that of $e \neq e_i$. By definition, $\getSet(v\cat z \cat \delete_{e_i})(e_i)= \getSet(v\cat z)(e_i) -1$. Since $z$ does not contain any operation $\delete_{e_i}$, $\getSet(v\cat z)(e_i)\geq \getSet(v)(e_i)$. Because $v\cat \delete_{e_i}\in L$ by assumption, we have $\getSet(v\cat \delete_{e_i})(e_i) = \getSet(v)(e_i)-1 \geq 0$, which implies with the earlier equality and inequality that $\getSet(v\cat z \cat \delete_{e_i})(e_i)\geq 0$. Considering now $e\neq e_i$, by definition $\getSet(v\cat z \cat \delete_{e_i})(e)= \getSet(v\cat z)(e)$, and since $\getSet(v\cat z)(e)\geq 0$ (as $v\cat z\in L$ by assumption), we also have $\getSet(v\cat z \cat \delete_{e_i})(e)\geq 0$.

\end{itemize}\vspace{-\baselineskip}~
\end{proofT}


\subsection{Shared Petri Net with Transition Rights}
Petri Nets are routinely used to model the production and consumption of resources, the synchronization of concurrent programs, or workflows. Informally, a Petri Net stores \emph{tokens} within \emph{places} and evolves through the firing of \emph{transitions}, which can be seen as predefined operations that consume tokens from some places and produce tokens in others.
To the best of our knowledge, no CRDT-based Petri Net has so far been proposed. One difficulty when distributing a Petri Net is ensuring that the number of tokens present in a place never becomes negative. Because two transitions might compete for the same tokens, they typically do not preserve each other's legality, breaking the commutativity assumption of CRDTs captured by~\cref{eq:CRDT:commutativity}. In the following, we overcome this challenge by ensuring that the rights to fire competing transactions (``transition rights'') are assigned to the same process. This design choice makes it possible to specify a Shared Petri Net with Transition Rights as a \nameOfObject.

\subsubsection{Petri Net} 
A Petri net is defined by the 5-tuple $(P, T, F, M_0, W)$ such as:
\begin{itemize}
	\item $P$ is a set of places;
	\item $T$ is a set of transitions ($P$ and $T$ are disjoints);
	\item $F \subseteq \big(P\times T\big) \cup \big(T\times P\big)$ is a set of directed edges from places to transitions or from transitions to places (thus creating a directed bipartite graph);
	\item $M_0 \in \mathbb{N}^{P}$ is an initial marking, that defines how many tokens each place initially contains;
	\item $W \in \left(\mathbb{N}_{>0}\right)^F$ indicates on each edge how many tokens are either consumed (for edges in $P\times T$) or produced (for edges in $T\times P$).
\end{itemize}

We note $\diamond$ the conflict relation between two transitions: $t_1\diamond t_2$ if $t_1$ and $t_2$ share an input place, or more formally if $N^-(t_1)\cap N^-(t_2)\neq\emptyset$, where $N^-$ denotes the direct predecessors of a vertex in the directed bipartite graph $(P\cup T,F)$. We note $\diamond^+$ the reflexive and transitive closure of $\diamond$. Since $\diamond$ is symmetric, $\diamond^+$ is an equivalence relation.

\newcommand{\fire}{{\sf fire}}
\newcommand{\getMarking}{{\sf marking}}

\subsubsection{Shared Petri Net with Transition Rights} 
A shared Petri net with transition rights is a shared object that provides the update operations $\fire_t$, which fires a given transition $t\in T$. To take into account conflicting transitions, the set of transitions is partitioned into a set $(T_i)_{p_i\in\Pi}$ of process-specific transitions, and a set of common transitions $T_C$, defined as follows:
\begin{itemize}
\item Two conflicting transitions belong to the same set $T_i$: for each $t\in T_i$, $[t]_{\diamond^+}\subseteq T_i$, where $[t]_{\diamond^+}$ is the equivalence class of $t$ for $\diamond^+$.
\item $T_C$ only contains transitions with no input places (i.e. transitions $t$ such that $N^-(t)=\emptyset$, and therefore $[t]_{\diamond^+}=\{t\}$).
\end{itemize}
A process $p_i$ can only fire transitions from $T_i\cup T_C$. 

A shared Petri net with transition rights supports a single query operation $\getMarking$ that returns the current marking $M\in\mathbb{Z}^{P}$ of the Petri net\footnote{As for multi-sets in \cref{sec:multi-set-with}, $\getMarking$ maps to $\mathbb{Z}$ so that it can be defined over $\mathbb{M}(T)$. The trace language $L$ defined below guarantees, however, by construction that $\getMarking$ remains positive.}. If $\getMarking$ is invoked after the $\fire$ operations contained in a trace $v$, then
\begin{equation}\label{eq:def:getMarking}
\getMarking(v)(p)\isdefinedas{}M_0(p) + \sum_{\substack{\fire_t\in v:\\t\in
N^-(p)}} W(t,p) - \sum_{\substack{\fire_t\in v:\\ t\in N^+(p)}}
W(p,t).
\end{equation}

\subsubsection{Alphabet and language definitions}
Formally, we define the sets $C$, $(O_i)_i$, and $Q$ of a shared Petri net with transition rights are as follows:
\begin{itemize}
	\item $C   \isdefinedas \{\fire_t\}_{t\in T_C}$,
  \item $O_i \isdefinedas \{\fire_t\}_{t\in T_i}$,
	\item $Q \isdefinedas \{\getMarking\}$.
\end{itemize}
$\fire_t$ are silent update operations, i.e. their $\outputFunc$ function always returns some pre-determined constant value (e.g. $\bot$).

The trace language of a shared Petri net is defined as
\begin{equation}
L \isdefinedas \big\{v\in \mathbb{M}(\Sigma) \;\big|\; \forall p\in P: \getMarking(v)(p)\geq 0\big\}.
\end{equation}

\begin{theorem}\label{th:multiPetri:PCO}
	$\big(C,(O_j)_j,L,Q\big)$ as defined above for a shared Petri net with transition rights specifies a \nameOfObject.
\end{theorem}

\begin{proofT}\renewcommand{\toto}{th:multiPetri:PCO}%
\begin{itemize}%
\item \emph{Initial emptiness:} By construction
$\getMarking(\epsilon)= M_0 \in \mathbb{N}^P$, hence $\epsilon\in L$.
\item \emph{$C^{\ast}$-Closure:} Consider a trace $v \in L$, and a fire operation $\fire_t\in C$, where $t\in T_C$. By $T_C$'s definition, $N^-(t)=\emptyset$, and therefore $\forall p\in P: t\not\in N^+(p)$. Injecting this property in \cref{eq:def:getMarking}, we have
\begin{align*}
\getMarking(v\cat\fire_t)(p)&=\getMarking(v) + \mathds{1}_{t\in N^-(p)}\times W(t,p) - \mathds{1}_{t\in N^+(p)}\times W(p,t),\\
&=\getMarking(v) + \mathds{1}_{t\in N^-(p)}\times W(t,p),\\
&\geq \getMarking(v).
\end{align*}
where $\mathds{1}_P$ represents the indicator function for predicate $P$.
\item \emph{$I$-diamond Closure:} Consider a trace $v \in L$, an operation $\fire_{t_i}$ with $t_i\in T_i$, and a trace of $\fire$ operations $z \in \mathbb{M}(\Sigma \setminus O_i)$, so that $\{v\cat \fire_{t_i}, v\cat z\}\subseteq L$. We have to consider two categories of places, depending on whether a place $p$ belongs to $N^-(t_i)$ or not. Case 1: For a place $p\not\in N^-(t_i)$, we have $t_i\not\in N^+(p)$, which injected in \cref{eq:def:getMarking} yields
\begin{align*}
\getMarking(v\cat z\cat \fire_{t_i})(p)
&= \getMarking(v\cat z)(p) + \mathds{1}_{t_i\in N^-(p)}\times W(t,p),\\
&\geq \getMarking(v\cat z)(p), \tag{since $W(t,p)>0$}\\
&\geq 0.\tag{since $v\cat z\in L$}
\end{align*}
Case 2: Consider a place $p\in N^-(t_i)$. For any update operations $\fire_{t'}\in z$, by definition of $T_i$ and since $z \in \mathbb{M}(\Sigma \setminus O_i)$, we have $\neg(t'\diamond^+ t_i)$, i.e. $t_i$ is not in conflict with any of the transitions $t'$ fired in the trace fragment $z$. As a result, $p\in N^-(t_i)$ implies $\forall\, \fire_{t'}\in z: p\not\in N^-(t')$, and equivalently $\forall\, \fire_{t'}\in z: t'\not\in N^+(p)$, and therefore
\begin{align*}
\getMarking(v\cat z)(p)
&= \getMarking(v)(p) + \textstyle\sum_{\substack{\fire_{t'}\in z:\\t'\in N^-(p)}} W(t',p),\\
&\geq \getMarking(v)(p).\tag{since $W(t',p)>0$}
\end{align*}
and therefore
\begin{align*}
\getMarking(v\cat z\cat \fire_{t_i})(p)
&= \getMarking(v\cat z)(p) + \mathds{1}_{t_i\in N^-(p)}\times W(t_i,p) - \mathds{1}_{t_i\in N^+(p)}\times W(p,t_i),\\
&\geq \getMarking(v)(p) + \mathds{1}_{t_i\in N^-(p)}\times W(t_i,p) - \mathds{1}_{t_i\in N^+(p)}\times W(p,t_i),\\
&\geq \getMarking(v\cat \fire_{t_i})(p),\\
&\geq 0,\tag{since $v\cat \fire_{t_i}\in L$}
\end{align*}
which yields $v\cat z\cat \fire_{t_i}\in L$ by definition of $L$.
\end{itemize}\vspace{-\baselineskip}~%
\end{proofT}

\subsection{Money Transfer}

A money transfer object formalizes the services provided by a cryptocurrency, for instance, in the context of a Blockchain. Following on the observation that a money transfer object has consensus number $1$ \cite{DBLP:journals/toplas/Herlihy91} (and hence does not require the total order typically provided by blockchains and related distributed ledger technologies), several lightweight algorithms have been proposed to implement money transfer in a Byzantine setting~\cite{auvolat:eatcs,DBLP:conf/aft/BaudetDS20,DBLP:conf/dsn/CollinsGKKMPPST20,DBLP:journals/dc/GuerraouiKMPS22}. In the following, we show that a money transfer object can be cast as a \nameOfObject.

\subsubsection{Definition}

A money transfer object~\cite{auvolat:eatcs,DBLP:journals/dc/GuerraouiKMPS22} is a distributed object that associates each process $p_i$ with an account, and two classes of operations: \emph{transfer} operations transfer some amount to another process' account, while \emph{balance} operations return the balance of a given process.

To demonstrate common updates, we also introduce (optional) \emph{mint} operations that create money and add it to a given account. In practice, such an operation would only be available to a limited set of trusted correct processes. In the following, however, we will assume for simplicity it is available to everyone w.l.o.g. Accounts cannot be negative, and the total amount of money in the system must be equal to the initial amount augmented by the value of each \emph{mint} operation.

\newcommand{\initial}{\mathit{init}}
\newcommand{\transfer}{\textsf{transfer}}
\newcommand{\balance}{\textsf{balance}}
\newcommand{\mint}{\textsf{mint}}
\newcommand{\plus}{\textsf{plus}}
\newcommand{\minus}{\textsf{minus}}
\newcommand{\acc}{\textsf{acc}}
\newcommand{\traceOp}{v}

\subsubsection{Notations}
\begin{itemize}
\item $\initial_i$ is the initial amount available in the
account of process $p_i$. For all $p_i$, $\initial_i\geq 0$.
\item $\transfer_{i,j,x}$ is the update operation \emph{transfer}
from $p_i$ to $p_j$ of an amount $x\geq 0$.
\item $\mint_{i,x}$ is the update operation \emph{mint}
of an amount $x \geq 0$ into $p_i$'s account.
\item For a process $p_i$ and a trace of update operations $\traceOp$, we define the following three functions:
\begin{align}
\plus(i,\traceOp)  \isdefinedas& \sum_{\transfer_{\_,i,y} \in \traceOp} y + \sum_{\mint_{i,y} \in \traceOp } y,\\
\minus(i,\traceOp) \isdefinedas& \sum_{\transfer_{i,\_,y} \in \traceOp} y,\\
\acc(i,\traceOp)   \isdefinedas& \initial_i + \plus(i,\traceOp) - \minus(i,v),
\end{align}
where $\textsf{x}\in v$ in the above formulas iterates over the elements of the trace $v$ that fit the pattern $\textsf{x}$.

	\item $\balance_{i}$ is the query operation \emph{balance} for a given account of a process $p_i$. If a
balance operation is invoked after the operations contained in a trace $v$, we have $\balance_{i}(v) = \acc(i,v)$.
\end{itemize}
$\transfer_{i,j,x}$ and $\mint_{i,x}$ are silent update operations, i.e. their respective $\outputFunc$ functions always return some pre-determined constant value (e.g. $\bot$).

\subsubsection{Alphabet and language definitions}
Formally, we define the sets of $C$, $(O_i)_i$, and $Q$ of a money transfer object as follows:
\begin{itemize}
	\item $C\isdefinedas \{\mint_{i,x}\}_{1\leq i \leq n, x\in \mathbb{R}_{\geq 0}}$,
  \item $O_i\isdefinedas \{\transfer_{i,j,x}\}_{1\leq j \leq n, x \in \mathbb{R}_{\geq 0}}$,
	\item $Q\isdefinedas \{\balance_i\}_{1 \leq i \leq n}$.
\end{itemize}

We define the trace language $L$ as
\begin{equation}
L \isdefinedas \big\{v\in \mathbb{M}(\Sigma) \;\big|\; \forall p_i \in \Pi: \acc(i,v)\geq 0\big\}.
\end{equation}

\begin{theorem}\label{th:money:transfer:PCO}
	$\big(C,(O_j)_j,L,Q\big)$ as defined above for a money-transfer object specifies a \nameOfObject.
\end{theorem}

\begin{proofT}\renewcommand{\toto}{th:money:transfer:PCO}
We need to prove that $L$ verifies the properties of \emph{Initial emptiness}, \emph{$C^{\ast}$-Closure}, and \emph{$I$-diamond Closure}.

\begin{itemize}

\item \emph{Initial emptiness:} It is easy to see that for any process $p_i$, $\plus(i,\epsilon)=\minus(i,\epsilon)=0$, and therefore that $\acc(i,\epsilon)= \initial_i$, where $\epsilon$ represents the empty trace. Since by hypothesis $\initial_i\geq 0$ (no process start with a negative balance), $\acc(i,\epsilon)\geq 0$, and $\epsilon \in L$, by definition of $L$.

\item \emph{$C^{\ast}$-Closure:} Consider a trace $v \in L$, and a money-creation operation on process $p_i$, $\mint_{i,x}\in C$. By definition of $v$, we have for all processes $p_j$ that $\acc(j,v)\geq 0$. By definition of $\plus()$, $\minus()$ and $\acc()$, we also have $\acc(i,v\cat \mint_{i,x})= \acc(i,v) + x$, and $\acc(j,v\cat \mint_{i,x})= \acc(j,v)$ for $j\neq i$. Combined with $x\geq 0$ and $\acc(j,v)\geq 0$ for all $p_j$, this implies that $\acc(j,v\cat \mint_{i,x})\geq 0$ for all processes $p_j$, and therefore that $v\cat \mint_{i,x}\in L$.

\item \emph{$I$-diamond Closure:} Consider a trace $v \in L$, a transfer $\transfer_{i,j,x}$, and a trace of updates $z \in \mathbb{M}(\Sigma \setminus O_i)$, so that $\{v\cat \transfer_{i,j,x}, v\cat z\}\subseteq L$. We have to consider the cases of $p_i$, $p_j$, and $p_\ell \not\in \{p_i,p_j\}$. \emph{Case of $p_i$}: By definition, $\acc(i,v \cat z \cat \transfer_{i,j,x}) = \acc(i,v\cat z) - x$, and $\acc(i,v \cat \transfer_{i,j,x}) = \acc(i,v) - x$. Because $z$ contains no transfer $\transfer_{i,-,-}$ from $p_i$, we have $\acc(i,v\cat z)\geq \acc(i,v)$, and therefore $\acc(i,v \cat z \cat \transfer_{i,j,x}) = \acc(i,v\cat z) - x \geq \acc(i,v) -x = \acc(i,v \cat \transfer_{i,j,x})$. Since $v \cat \transfer_{i,j,x}\in L$, $\acc(i,v \cat \transfer_{i,j,x})\geq 0$, which implies that $\acc(i,v \cat z \cat \transfer_{i,j,x})\geq 0$. \emph{Case of $p_j$}: By definition $\acc(j,v \cat z \cat \transfer_{i,j,x})= \acc(j,v \cat z) + x$. With $\acc(j,v \cat z)\geq 0$ (since $v \cat z \in L$), and $x\geq 0$ (by definition), this yields $\acc(j,v \cat z \cat \transfer_{i,j,x})\geq 0$. \emph{Case of $p_\ell \not\in \{p_i,p_j\}$}: $\acc(\ell,v \cat z \cat \transfer_{i,j,x})= \acc(\ell,v \cat z)$, which implies $\acc(\ell,v \cat z \cat \transfer_{i,j,x})\geq 0$ since $v \cat z\in L$. These three cases show that $\acc(k,v \cat z \cat \transfer_{i,j,x})\geq 0$ for all processes $p_k$, and therefore that $v \cat z \cat \transfer_{i,j,x} \in L$.
  
\end{itemize}\vspace{-\baselineskip}~
\end{proofT}

\subsection{Idempotent Work Stealing Deque}

\newcommand{\WSD}{WSD\xspace}

Work stealing is a classical approach to implement load-balancing within a parallel computing infrastructure with irregular workloads (e.g., a massively parallel computer or a computing grid)~\cite{Blumofe:WorkStealing:10.1145/324133.324234,hendler2004dynamic,michael2009idempotent,DBLP:conf/sc/DinanLSKN09}. In keeping with the vocabulary used so far, we shall consider a system in which $n$ processes submit tasks. A process keeps the tasks it has submitted but which have not been executed yet in a semi-private data structure called a \emph{work stealing deque} (\WSD for short, more on this below), and executes in priority its own tasks (termed local tasks). When a process no longer has any local task to execute (i.e., when its local deque is empty), it steals a pending task from some other process, selected at random.

Because we consider fault-models in which processes might crash or behave in a Byzantine manner, we consider the case of \emph{idempotent work stealing}~\cite{michael2009idempotent}. In this relaxed version of work stealing, every task submitted by (correct) processes must eventually be executed at least once, but might be executed several times (contrary to exactly once in traditional work-stealing~\cite{DBLP:conf/popl/AttiyaGHKMV11}).

\subsubsection{A Multiple Idempotent Work Stealing Deque}
\label{sec:work-stealing-deque}
Work stealing deques, or \WSD{}s, lie at the heart of any work stealing algorithm. A \WSD typically supports local LIFO \emph{push} and \emph{pop} operations~\cite{hendler2004dynamic}, and allows remote processes to perform FIFO pop operations when they attempt to ``steal'' work from other processes. Because some processes might fail in our model, the theft of a task from a correct process by a faulty one could lead to the loss of this task. Our implementation, therefore, slightly changes the default semantic of \WSD{}s and implements the steal operation as a query operation that returns the top task of a deque, but does not modify the deque. Thief processes notify the owner of a task that the task is completed simply by publishing its result. The removal proper of the task from the local deque is performed by the owning process. To tolerate Byzantine behavior, the proposed approach assumes that correct processes can check that the result of a task is indeed valid, e.g., by applying some predicate on the return result. (We return to this last point at the end of this section.)

A work-stealing algorithm uses as many deques as there are processes. For simplicity, we specify the overall behavior of this set of deques, which we call a \emph{Multiple Idempotent Work Stealing Deque}.

\newcommand{\pushBottom}{{\sf pushBottom}}
\newcommand{\popBottom}{{\sf popBottom}}
\newcommand{\remove}{{\sf remove}}
\newcommand{\addResult}{{\sf addResult}}
\newcommand{\steal}{{\sf getTop}}
\newcommand{\getBot}{{\sf getBottom}}
\newcommand{\getPend}{{\sf getPending}}
\newcommand{\getResults}{{\sf getResults}}
\newcommand{\set}{{\sf set}}

\newcommand{\state}{{\sf d\_state}}

Formally, we define a \emph{Multiple Idempotent Work Stealing Deque} as a distributed object which provides four update operations (\pushBottom, \popBottom, \remove, and \addResult) and four query operations (\steal, \getBot, \getPend, and \getResults):
\begin{itemize}
\item $\pushBottom_{i,t}$ adds (pushes) task $t$ to the bottom of the deque of processor $p_i$. Only $p_i$ might execute this operation.
\item $\popBottom_{i}$ removes (pops) and returns the task at the bottom of the deque of processor $p_i$ in LIFO order. Only $p_i$ might execute this operation.
\item $\remove_{i,t}$ removes task $t$ from the deque of processor $p_i$ (independently of its position). If $t$ is not in $p_i$'s deque, it has no effect.
\item $\addResult_{t,r}$ records that the result of task $t$ is some value $r$. $\addResult_{t,r}$ is used by thief processes to notify the task owner of $t$ that a (valid) result is available for $t$.
\end{itemize}
The following operations are query operations, i.e., they do not modify the content of the deque.
\begin{itemize}
\item $\steal_{i}$ returns the task at the top of $p_i$'s deque in FIFO order.
\item $\getBot_{i}$ returns the task at the bottom of $p_i$'s deque in LIFO order/
\item $\getPend_{i}$ returns the set of pending tasks, i.e., the task currently stored in $p_i$'s deque.
\item $\getResults_{t}$ returns the set of results published for task $t$ by thief processes (i.e. by processes different from $t$'s owner) using $\addResult_{t,-}$.
\end{itemize}
Only $p_i$ might execute $\pushBottom_{i,t}$, $\popBottom_{i}$, or $\remove_{i,t}$. $\addResult_{t,r}$ can be executed by any process, but in a Byzantine context the process that submitted $t$ must be able to check the validity of the result $r$. $\steal_{i}$, $\getBot_{i}$, $\getPend_{i}$, and $\getResults_{t}$ are query operations available to any process. We assume tasks have unique IDs that can only be used once, and that IDs are partitioned between processes (e.g., by incorporating process IDs) so that a process $p$ cannot submit task using $\pushBottom_{i,-}$ using an ID allocated to another process.

\paragraph{Alphabet and Notations}
\label{sec:alphabet-notations}

The sets $C$, $(O_i)_i$, and $Q$ of a Multiple Idempotent
Work Stealing Deque are defined as
\begin{itemize}
	\item $C_{\phantom{i}} \isdefinedas \{\addResult_{t,r}\mid r \text{ is a valid result for task } t\}_{(t,r)\in\mathbb{T}\times\mathbb{V}}$,
  \item $O_i             \isdefinedas \{\pushBottom_{i,t},\remove_{i,t}\}_{(p_i,t)\in\Pi\times\mathbb{T}_i}\cup\{\popBottom_{i}\}_{p_i\in\Pi}$,
	\item $Q_{\phantom{i}} \isdefinedas \{\steal_{i},\getBot_i,\getPend_{i}\}_{p_i\in\Pi}\cup\{\getResults_t\}_{t\in\mathbb{T}}$,
\end{itemize}
where $\mathbb{T}$ is the set of all possible tasks, $\mathbb{T}_i$ is the set of tasks that $p_i$ may submit ($\mathbb{T}=\bigcup_i \mathbb{T}_i$), and $\mathbb{V}$ the set of task results.

\paragraph{The $\state$ helper function}

To help with the definition of the trace language $L$, we introduce a helper function, called $\state(i,v)$, that returns the current content of $p_i$'s deque following the operations recorded in a trace $v$.
\begin{equation*}
\begin{array}{cccc}
\state:&[1..n]\times\mathbb{M}(\Sigma)&\to&\mathbb{T}^{\ast}\cup \{\bot\}\\
&(i,v)&\mapsto& \state(i,v).
\end{array}
\end{equation*}

In what follows, $s\text{\textbackslash} t$ is the sequence obtained when removing all occurrences of the task $t\in\mathbb{T}$ from the sequence of tasks $s\in\mathbb{T}^{\ast}$. $s[i..j]$ represents the subsequence of $s$ containing its $i^{\text{th}}$ to $j^{\text{th}}$ elements\footnote{Said differently, if $s=(s_k)_k$, $s[i..j]=(s_k)_{i\leq k \leq j}$.}. We define the special case in which $i>j$ as follows:
\begin{align}
s[i..i-1]&=\epsilon,\\
s[i..j]&=\bot\quad\text{ if }i>j+1.\label{eq:s:i:j:bot}
\end{align}
where $\epsilon$ is the empty sequence, and $\bot$ is a non-value.

Using these notations, $\state(i,v)$ is defined recursively as follows.
\begin{align}
\state(i,\epsilon)&\isdefinedas\epsilon,\label{eq:state:epsilon}\\
\intertext{\hspace{3em}if $\state(i,v)\neq\bot$ then}
\state(i,v\cat \pushBottom_{i,t})&\isdefinedas\state(i,v)\cat t,\\
\state(i,v\cat \popBottom_{i})\phantom{!_{,t}}&\isdefinedas\state(i,v)\big[2..|\state(i,v)|\big],\label{eq:opBot:def}\\
\state(i,v\cat \remove_{i,t})\phantom{ttom}&\isdefinedas\state(i,v)\text{\textbackslash} t,\\
\intertext{\hspace{3em}if $\state(i,v)=\bot$ then}
\state(i,v\cat \op)&\isdefinedas\bot.\label{eq:def:state:if:bot}
\end{align}
As a result, $\state(i,v)$ switches to the non-value $\bot$ when a $\popBottom$ operation is applied to an empty deque in \cref{eq:opBot:def}. Once $\state(i,v)$ has collapsed to $\bot$, \cref{eq:def:state:if:bot} ensures it keeps this value. Other update operations leave $\state(i,v)$ unchanged.

\begin{lemma}\label{lemma:state:well:defined}
$\state$ is a well-defined function on $\mathbb{M}(\Sigma)$.
\end{lemma}
\begin{proofL}
  Because a trace $v$ only imposes a partial order on its operations, the above recursive definitions could be ambiguous. In the case of $\state$, however, because the operations of $O_i$ do not commute, a trace $v$ imposes a total order on the operations $\pushBottom_{i,t}$, $\popBottom_{i}$, and $\remove_{i,t}$, insuring that \state{} is well defined.
  \renewcommand{\toto}{lemma:state:well:defined}%
\end{proofL}

\paragraph{Language Definition}

The trace language of a Multiple Idempotent Work Stealing Deque is defined as
\begin{equation*}
L \isdefinedas \big\{v\in \mathbb{M}(\Sigma) \mid \forall p_i\in\Pi, t\in \mathbb{T}: |\proj{v}{\pushBottom_{i,t}}|\leq 1 \wedge \state(i,v)\neq \bot \big\},
\end{equation*}
which essentially states that $p_i$ should only submit a task $t$ at most once, and that $\popBottom_{i}$ is forbidden when $p_i$'s deque is empty, due to the recursive definition of \cref{eq:opBot:def}, and the fact that $\state(i,v)\big[2..0\big]=\bot$ by definition.

\paragraph{Defining the Query and Result Functions}

Based on \state, the query functions $\steal_{i}$, $\getBot_{i}$ and
$\getPend_{i}$ are defined on $L$ as
\begin{align*}
\steal_{i}(v)     &\isdefinedas \state(i,v)\big[|\state(i,v)|\big],\\
\getBot_{i}(v)    &\isdefinedas \state(i,v)\big[1\big], \\
\getPend_{i}(v)   &\isdefinedas \set(\state(i,v)),\\
\getResults_{t}(v)&\isdefinedas \{r\in\mathbb{V}\mid \addResult_{t,r}\text{ appears at least once in } v\},
\end{align*}
where $s[i]$ is the $i^{\text{th}}$ element of sequence $s$.
Similarly when $\popBottom_{i}$ is appended to a trace $v$, it returns the value $\state(i,v)[1]$:
\begin{equation*}
\outputFunc(\popBottom_{i},v)\isdefinedas\state(i,v)[1].
\end{equation*}
The update operations $\pushBottom_{i,t}$, $\remove_{i,t}$, and $\addResult_{t,r}$ are silent.

In the above definitions, by convention, we will assume $s[i]=\bot$ when $i\not\in[1..|s|]$. As a result, $\steal_{i}(v)$, $\getBot_{i}(v)$, and $\outputFunc(\popBottom_{i},v)$ returns a non-value ($\bot$) when $p_i$'s deque is empty.

\begin{lemma}\label{lemma:query:output:update:well:defined}
$\steal_{i}$, $\getBot_{i}$, $\getPend_{i}$, $\getResults_{t}$ and $\outputFunc(\popBottom_{i},-)$ are well-defined over $L$. Futhermore, for all $v\in L$, if $v\cat \popBottom_{i}\in L$, then $\outputFunc(\popBottom_{i},v)\neq\bot$.
\end{lemma}

\begin{proofL}
Lemma~\ref{lemma:state:well:defined} implies that $\steal_{i}$,
$\getBot_{i}$, $\getPend_{i}$ and $\outputFunc(\popBottom_{i},-)$
are well defined. $\getResults_{t}$ is well defined by
construction of $\mathbb{M}(\Sigma)$.

Consider $v\in L$, such that $v\cat \popBottom_{i}\in L$. By definition of $L$,
\begin{align}
\state(i,v)&\neq\bot,\label{eq:state:v:neq:bot}\\
\state(i,v \cat \popBottom_{i})&\neq\bot.\label{eq:state:cat:popBottom:neq:bot}
\end{align}
By definition of $\state$, \cref{eq:state:v:neq:bot} implies that $\state(i,v \cat \popBottom_{i})$ was obtained through \cref{eq:opBot:def}. The only case in which the range operator applied on a sequence might return $\bot$ is defined by
\cref{eq:s:i:j:bot}. As a result, \cref{eq:state:cat:popBottom:neq:bot} implies that $|\state(i,v)|\geq 1$, which yields $\outputFunc(\popBottom_{i},v)\neq\bot$ by definition of $\outputFunc(\popBottom_{i},v)$.
\renewcommand{\toto}{lemma:query:output:update:well:defined}%
\end{proofL}

\begin{theorem}\label{th:multiple:work:stealing:PCO}
	$\big(C,(O_j)_j,L,Q\big)$ as defined above for a Multiple Idempotent Work Stealing Deque specifies a \nameOfObject.
\end{theorem}

\begin{proofT}\renewcommand{\toto}{th:multiple:work:stealing:PCO}
We need to prove that $L$ verifies the properties of \emph{Initial emptiness}, \emph{$C^{\ast}$-Closure}, and \emph{$I$-diamond Closure}.

\begin{itemize}
\item \emph{Initial emptiness} follows from the definition of $L$ and \cref{eq:state:epsilon}.
  
\item \emph{$C^{\ast}$-Closure} follows from the fact that $\addResult_{t,r}$ does not modify the value of $\state(i,-)$, i.e. $$\state(i,v\cat \addResult_{t,r}) = \state(i,v).$$

\item \emph{$I$-diamond Closure:} Consider $v \in L$, an update $\op_i \in O_i$ $=$ $\{\pushBottom_{i,t},$ $\popBottom_{i},$ $\remove_{i,t}\}_{(p_i,t)\in\Pi\times\mathbb{T}}$ , and some trace of updates $z \in \mathbb{M}(\Sigma \setminus O_i)$ that are independent of $O_i$. Assume $\{v\cat \op_i, v\cat z\}\subseteq L$. As only the operations of $O_i$ modify the value of $\state(i,-)$, all operations of $z$ leave $\state(i,v)$ unchanged, we therefore have by definition of $\state$
\begin{align*}
\state(i,v\cat z)&=\state(i,v),\\
\state(i,v\cat z \cat op_i)&=\state(i,v \cat op_i).
\end{align*}
As $v\cat \op_i\in L$, $\state(i,v \cat op_i)\neq\bot$ by definition of $L$, which with the above equality implies
\begin{equation}\label{eq:v:z:opi:neq:bot}
\state(i,v\cat z \cat op_i)\neq\bot.
\end{equation}     
By a similar reasoning, $\pushBottom_{i,t}\not\in z$ leads to $|\proj{v}{\pushBottom_{i,t}}|=|\proj{v\cat z}{\pushBottom_{i,t}}|$. This equality and the fact that $v\cat \op_i\in L$ yield in turn
\begin{equation}\label{eq:proj:v:z:opi:leq:1}
|\proj{v\cat z \cat \op_i}{\pushBottom_{i,t}}|=|\proj{v\cat \op_i}{\pushBottom_{i,t}}| \leq 1.
\end{equation}
\cref{eq:v:z:opi:neq:bot} and \cref{eq:proj:v:z:opi:leq:1} mean that $v\cat z \cat \op_i\in L$ by definition of $L$, concluding the proof.
\end{itemize}\vspace{-\baselineskip}~
\end{proofT}

\newcommand{\submitTask}{{\sf submit\_task}}
\newcommand{\execute}{{\sf execute}}
\newcommand{\publishResult}{{\sf publish\_result}}

\subsubsection{Using a Multiple Idempotent Work Stealing Deque}

Algorithm~\ref{alg:work-stealing:example} illustrates how the Multiple Idempotent Work Stealing Deque specified in Section~\ref{sec:work-stealing-deque} could be used. The algorithm provides the local function $\submitTask()$, and uses the callback $\publishResult()$ (lines~\ref{line:publishResult:1} and~\ref{line:publishResult:2}) to notify the application that a task has been completed. $\submitTask(t)$ simply pushes the task $t$ onto the bottom of $p_i$'s dequeue object (\cref{line:pushBot:i:t}). The rest of the algorithm is implemented through three event-based handlers:
\begin{itemize}

\item The event handler at lines~\ref{line:EH:local:tasks:start}-\ref{line:EH:local:tasks:end} executes the task found in $p_i$'s deque, i.e. the task submitted locally by $p_i$ using $\submitTask$. Once completed, the task's result is passed on to the application using the callback $\publishResult$.

\item The event handler at lines~\ref{line:EH:task:stealing:start}-\ref{line:EH:task:stealing:end} triggers when $p_i$'s deque becomes empty. $p_i$ selects at random another process $p_j$ whose deque still contains some task(s), and steals one of $p_j$'s tasks to execute it.

\item The event handler at lines~\ref{line:EH:stolen:task:report:result:start}-\ref{line:EH:stolen:task:report:result:end} reacts to the publication of results by remote processes for pending local tasks (i.e., local tasks that are still in $p_i$'s deque). When a result for such a task is found, the task is removed from $p_i$'s deque, and its result is passed on to the application through the callback $\publishResult$.

\end{itemize}

\begin{algorithm}[tb]
\Function(\Comment*[f]{$t\in\mathbb{T}_i$}){$\submitTask(t)$}{\label{line:submitTask}
$\applyFunc(\pushBottom_{i,t})$\;\label{line:pushBot:i:t}
}
\AlgoSkip
\WhenDo{$\applyFunc(\getBot_i)\neq\bot$\label{line:EH:local:tasks:start}}{
$t \leftarrow \applyFunc(\popBottom_{i})$\label{line:popBottom:i}\;
$r \leftarrow \execute(t)$\label{line:execute:local}\;
$\publishResult(t,r)$\;\label{line:publishResult:1}\label{line:EH:local:tasks:end}
}
\AlgoSkip
\WhenDo{$\applyFunc(\getBot_i)=\bot \wedge \Exists p_j\neq p_i: \applyFunc(\steal_j)\neq\bot $%
\label{line:EH:task:stealing:start}}{
$t \leftarrow \applyFunc(\steal_j)$\;
$r \leftarrow \execute(t)$\label{line:execute:remote}\;
$\applyFunc(\addResult_{t,r})$\;\label{line:EH:task:stealing:end}
}
\AlgoSkip
\WhenDo{$\Exists t\in\applyFunc(\getPend_i): \applyFunc(\getResults_t)\neq\emptyset$%
\label{line:EH:stolen:task:report:result:start}}{
$\applyFunc(\remove_{i,t})$\label{line:remove:i:t}\;
$r \leftarrow \text{choose some value from }\applyFunc(\getResults_t)$\label{line:choose:r}\;
$\publishResult(t,r)$\;\label{line:publishResult:2}\label{line:EH:stolen:task:report:result:end}
}
\caption{A distributed work stealing algorithm using the Multiple Idempotent Work Stealing Deque specified as a Process-Commutative Object in Section~\ref{sec:work-stealing-deque}. (Code for $p_i$.)}\label{alg:work-stealing:example}
\end{algorithm}

Algorithm~\ref{alg:work-stealing:example} assumes that a correct process $p_i$ only submits tasks with IDs taken from $\mathbb{T}_i$, and never submits any task $t\in\mathbb{T}_i$ more than once. (If any of these two conditions were not respected, \cref{line:pushBot:i:t} would block due to the definition of $\Sigma$, $L$, and the condition at \cref{line:apply:update:test} of Algorithm~\ref{algo:generic}.) Algorithm~\ref{alg:work-stealing:example} further assumes that $\submitTask()$ and each of the event handlers is executed in mutual exclusion (equivalently are placed in the same monitor), using a fair scheduler, in the sense that any of these four blocks of code cannot remain indefinitely activated without being eventually executed. (We discuss below how these concurrency assumptions can be relaxed.)

Under the following assumptions, Algorithm~\ref{alg:work-stealing:example} ensures that if a correct process stops submitting tasks, then, eventually, all tasks submitted by this correct process are executed, and a valid result is returned to the application exactly once for each of them, as captured by the following theorem.

\begin{theorem}\label{theo:finitesubmission:alltask:executed}
Using Algorithm~\ref{alg:work-stealing:example} either in the $\CAMP$ or $\BAMP$ models, if a correct process $p_i$ invokes $\submitTask(-)$ a finite number of times, then each task submitted by $p_i$ is eventually executed, and a valid result for each task is returned exactly once to the application using the callback $\publishResult(-,-)$.
\end{theorem}

\begin{proofT} Consider an execution of Algorithm~\ref{alg:work-stealing:example}, and $p_i$, a correct process. Let us note $H=(E_1,\dots, E_n)$ the history recording the invocations of operations on the underlying Multiple Idempotent Work Stealing Deque object. Assuming a crash (resp. Byzantine) failure model, let us note $H'=(E_1,\dots, E_n)$ the crash-corrected history (resp. mock history) corresponding to $H$ (Section~\ref{sec:distr-name}). Because $p_i$ is correct, we have $E'_i=E_i$. As earlier, let us note $\widehat{H'}_i = (\widehat{E'}_1, \dots, \widehat{E'}_n)$ $p_i$'s \emph{perception} of $H'$. By definition, $\widehat{E'}_i=E'_i=E_i$.

Because of Theorem~\ref{th:multiple:work:stealing:PCO}, we can apply Theorem~\ref{crash-proof} (resp. Theorem~\ref{byzantine-proof}), which yields the existence of a \nameOfObjectAbbr-legal serialization $S_i$ of $\widehat{H'}_i$. $S_i$ contains all the update operations of $H'$, but only the queries of $E_i$.

Assume $p_i$ only invokes $\submitTask(\cdot)$ a finite number of times. Further assume some task $\tau\in\mathbb{T}_i$ submitted by $p_i$ is never returned by $\popBottom_{i}$ at \cref{line:popBottom:i} or removed by $\remove_{i,\tau}$ at \cref{line:remove:i:t} in $S_i$. Using Query Validity (Section~\ref{sec:sequ-spec}), and the definitions of $\pushBottom_{i}$, $\popBottom_{i}$ and $\remove_{i,\tau}$, this means that for any prefix $s$ of $S_i$ that contains the $\pushBottom_{i,\tau}$ operation applied at \cref{line:pushBot:i:t}, we have $\tau\in\state(s,i)$. This implies that any $\getBot_i$ query occurring after $\pushBottom_{i,\tau}$ cannot return $\bot$ (let A be this property). As a result, the event handler at lines~\ref{line:EH:task:stealing:start}-\ref{line:EH:task:stealing:end} of Algorithm~\ref{alg:work-stealing:example} will remain activated continuously after $\pushBottom_{i,\tau}$. By scheduling assumption, this implies that \cref{line:popBottom:i} will be executed arbitrarily often.

Because $p_i$ only invokes $\submitTask(\cdot)$ a finite number of times, an arbitrary number of $\popBottom_{i}$ operations must eventually empty $\state(s,i)$, leading to $\getBot_i$ returning $\bot$ at some point, thus contradicting the property A stated above. We conclude that $\tau$ must be returned by $\popBottom_{i}$ at \cref{line:popBottom:i} or removed by $\remove_{i,t}$ at \cref{line:remove:i:t} in $S_i$.

If $\tau$ is returned by $\popBottom_{i}$ at \cref{line:popBottom:i}, because $p_i$ is correct, it executes the rest of the event handler and returns a valid result to the application using $\publishResult$. If $\tau$ is removed at \cref{line:remove:i:t}, by definition of $\getResults_t$ and the construction of the set $C$ (which only allow for valid results to be included in traces of $L$), the result $r$ found at \cref{line:choose:r} is also valid, and $p_i$ returns it to the application.

The fact that $\pushBottom_{i,\tau}$ can only be included once in $S_i$, and that $\popBottom_{i}$ (when it returns $\tau$ at \cref{line:popBottom:i}) and $\remove_{i,\tau}$ (at \cref{line:remove:i:t}) both remove $\tau$ from $\state(s,i)$ insure that $\publishResult$ can only be invoked at most once for $\tau$ under our concurrency assumption of mutual exclusion of the event handlers.
\renewcommand{\toto}{theo:finitesubmission:alltask:executed}%
\end{proofT}

\paragraph{Relaxing the concurrency assumptions}
The result of \cref{theo:finitesubmission:alltask:executed} continues to hold if one relaxes the concurrency assumptions slightly, by releasing the monitor of \cref{alg:work-stealing:example} when calling $\execute(t)$ at \cref{line:execute:local}, and~\ref{line:execute:remote}, while maintaining a mutual exclusion between the first (\crefrange{line:EH:local:tasks:start}{line:EH:local:tasks:end}) and the second event handler (\crefrange{line:EH:task:stealing:start}{line:EH:task:stealing:end}). This approach provides more concurrency as it allows the application to submit tasks at \cref{line:submitTask,line:pushBot:i:t} and remote results to be published at \cref{line:publishResult:2} even if some task is already executing locally.

\paragraph{Relaxing the need to check the validity of returned results}
\label{sec:relaxing-need-check}
To work in a Byzantine setting, \cref{alg:work-stealing:example} assumes that correct processes can check whether a result $r$ is valid using some application-specific predicate. The assumption is tucked into the definition of the set $C$ in \cref{sec:alphabet-notations} (page~\pageref{sec:alphabet-notations}), which implies that $r$ must be valid for the update $\addResult_{t,r}$ to be considered an element $C$. If the multiple deque is implemented using \cref{algo:generic}, this verification occurs when testing the condition $\update \in C$ at \cref{line:rdelivery:test} on page~\pageref{algo:generic}.

When such a predicate is not available, an alternative approach can be to modify \crefrange{line:EH:stolen:task:report:result:start}{line:EH:stolen:task:report:result:end} of \cref{alg:work-stealing:example} so that a result is only accepted if it has been returned by enough processes (typically $t+1$, where $t$ is the maximum number of Byzantine processes in the system), a mechanism commonly used for instance in voluntary computing platforms.

\section{Conclusion}
\label{sec:conclusion}

This paper introduced {\it \nameOfObjects} (\nameOfObjectAbbrs), a class of distributed data structures that covers a wide range of practical use cases, from money-transfer systems to concurrent work-stealing deques. \nameOfObjectAbbrs are ``local-first'' data-structures, and thus similar to CRDTs (\emph{Conflict-free Replicated Data Types}). They go further than CRDTs by handling conflicts between operations issued by the same participant and offer a stronger consistency level that combines Strong Eventually Consistency (SEC) and Pipeline Consistency (PC). Computationally, \nameOfObjectAbbrs are weaker than atomic shared registers and can be implemented using solely a reliable broadcast primitive, both in a Byzantine and Crash failure model.

The paper formalized \nameOfObjectAbbrs using Mazurkiewicz traces and introduced a simple generic algorithm to implement \nameOfObjectAbbrs. The proposed algorithm is generic on two counts: (i) it can implement any \nameOfObjectAbbr (it is thus universal for \nameOfObjectAbbr objects), and (ii) it inherits the fault-tolerance assumptions of its underlying broadcast (it can be made crash- or Byzantine tolerant simply by using a crash- resp. Byzantine tolerant reliable FIFO broadcast).

To illustrate the interest of \nameOfObjectAbbrs, this work finally presented four practical applications of \nameOfObjectAbbrs, namely, money transfer, multi-sets with deletion rights, Petri nets, and work-stealing deques. 

\section*{Funding and Competing interests}

The authors state that they have no competing interests to declare.
This work was partially supported by the French \emph{Agence Nationale de la Recherche} (ANR) under the projects ByBloS (ANR-20-CE25-0002-01) and PriCLeSS (ANR-10-LABX-07-81, CominLabs Laboratory of excellence) devoted to the modular design of building blocks for large-scale Byzantine-tolerant multi-users applications.

\printbibliography

\newpage
\appendix

\section{Converting a \nameOfObjectAbbr specification to a state automaton}
\label{app:conv-name-spec}

As with word languages, trace languages are closely related to  automata~\cite{DBLP:reference/parallel/DiekertM11,DBLP:conf/ershov/Zielonka89,DBLP:journals/acta/DiekertM94,DBLP:books/daglib/0077825}. In the following, we exploit this relationship to adapt \cref{algo:generic} so that it uses an automaton rather than a trace language.

A \nameOfObjectAbbr specification $\big(C,(O_j)_{j\in[1..n]},L,Q\big)$ can be converted to a state automaton by exploiting the following equivalent relation $\simeq$ on traces of $L$.
\begin{equation}
  \forall u,v\in L: u \simeq v \text{ iff } \left\{
  \begin{array}{@{}l}
    \{w\in\mathbb{M}(\Sigma,I) \mid u\cat w \in L\} = \{w\in\mathbb{M}(\Sigma,I) \mid v\cat w \in L\} \:\wedge\\
    \forall w\in \{w'\in\mathbb{M}(\Sigma,I) \mid u\cat w' \in L\}:\\
    \quad\left|\begin{array}{@{\hspace{0.25em}}l}
    \forall \update\in \Sigma: \outputFunc(\update,u\cat w) = \outputFunc(\update,v\cat w) \:\wedge\\
    \forall q\in Q: q(u\cat w) = q(v\cat w).
    \end{array}\right.
  \end{array}\right.
\end{equation}
Two traces $u$ and $v$ in $L$ are equivalent according to $\simeq$ if they have indistinguishable futures, i.e., if the sequence of updates they allow are the same (first set equality), and if the `states' obtained after applying the same sequence $w$ to $u$ and $v$ produce the same outputs and query results.

An automaton can be constructed based on $ \simeq$ by taking the quotient set of $L$ by $\simeq$,
\begin{equation}
 S = \faktor{L}{\simeq},
\end{equation}
and by defining a state-transition function based on the concatenation operator $\cat$,
\begin{align}
  \delta: (S\times\Sigma) &\to \hspace{2em}S\\
  ([u]_{\simeq},\update)&\mapsto 
  \left\{\begin{array}{@{\hspace{0.1em}}ll}
    [u\cat\update]_{\simeq}&\text{if } u\cat\update\in L\\
    \text{\emph{undefined}}&\text{otherwise}.
  \end{array}\right.
\end{align}
Query and output functions naturally translate to $S$ using $q([u]_{\simeq})=q(u)$. The initial state of the automaton is simply the equivalence class of the empty chain, $s_0=[\epsilon]_\simeq$.

\cref{algo:generic} (\cref{sec:algorithm}) can then be adapted to use the automaton $(S,\delta,s_o)$ by modifying the following four lines.
\let\oldnl\nl
\newcommand{\nonl}{\renewcommand{\nl}{\let\nl\oldnl}}
\newcommand{\verticalDots}{\nonl\quad$\vdots$}
\begin{algorithm}[h!]
  \KwInit{$\statevar_i \leftarrow s_0$ ; ...}

  \verticalDots\;

  \setcounter{AlgoLine}{\value{value:apply:update:test}}
  \lIf(\Comment*[f]{\update{} must be authorized to $p_i$ and legal}){$\update \in C \cup O_i \wedge \delta(\statevar_i, \update)$ is defined}{}

  \verticalDots\;

  \setcounter{AlgoLine}{\value{value:deliver:update:seqi}}
  \wwait{$(\update \in C \cup O_j \wedge\sn=del_i[j]+1 \wedge \delta(\statevar_i, \update)$ is defined$)$}\;
$\statevar_i \leftarrow \delta(\statevar_i, \update)$\;
\end{algorithm}

Coming back to the toy token ring example of \cref{sec:few-remarks-diff}, the language $L = \big\{u\in \mathbb{M}(\Sigma,I) \bigm| 0 \leq \multip{u}(\tab)-\multip{u}(\tba)\leq 1 \big\}$ admits two equivalence classes under $\simeq$,
\begin{align}
  \faktor{L}{\simeq} &= \{L_A, L_B\}, \text{ with}\\
  L_A&= \big\{u\in L \bigm| \multip{u}(\tab)-\multip{u}(\tba)=0 \big\}, \text{ and}\\
  L_B&= \big\{u\in L \bigm| \multip{u}(\tab)-\multip{u}(\tba)=1 \big\}.
\end{align}
The resulting automaton has thus two states ($L_A$ when the token is with Alice, and $L_B$ when it is with Bob) and starts in $s_0=L_A$. The update $t_{AB}$ is valid in $L_A$ and transitions the automaton to $L_B$. $t_{BA}$ performs the opposite action, from $L_B$ to $L_A$.

\end{document}